\documentclass[aps,prd,11pt, tightenlines, oneside, notitlepage, superscriptaddress, showpacs, preprintnumbers, nofootinbib, onecolumn, fleqn, longbibliography, a4paper, floatfix]{revtex4-1}
\pdfoutput=1
\usepackage[centering,hmargin=2.5cm,vmargin=2.5cm]{geometry}
\usepackage{hyperref}
\usepackage[utf8]{inputenc}
\usepackage[normalem]{ulem}
\usepackage{amsmath,amssymb}
\usepackage{pifont}
\usepackage{mathtools}  
\usepackage{epsfig}
\usepackage{afterpage}
\usepackage{graphicx} 
\usepackage{url}
\usepackage{color}
\usepackage{slashed}
\usepackage{multirow}
\usepackage{placeins}
\usepackage{csquotes}
\usepackage{enumitem}
\usepackage[dvipsnames]{xcolor}
\usepackage{epstopdf}
\usepackage{soul}
\usepackage[capitalise, english]{cleveref}
\usepackage[separate-uncertainty=true]{siunitx}
\usepackage{mathrsfs}
\usepackage{xspace}

\graphicspath{{plots/}}

\hypersetup{
        pdfnewwindow=true,   
        colorlinks=true,     
        linkcolor=blue,      
        citecolor=blue,      
        filecolor=blue,      
        urlcolor=blue        
}

\usepackage{tikz} 
\usetikzlibrary{trees}
\usetikzlibrary{decorations.pathmorphing}
\usetikzlibrary{decorations.markings}

\DeclareSIUnit\year{yr}
\DeclareSIUnit\years{yrs}
\DeclareSIUnit\pc{pc}

\usepackage{array}
\newcolumntype{L}[1]{>{\raggedright\let\newline\\\arraybackslash\hspace{0pt}}m{#1}}
\newcolumntype{C}[1]{>{\centering\let\newline\\\arraybackslash\hspace{0pt}}m{#1}}
\newcolumntype{R}[1]{>{\raggedleft\let\newline\\\arraybackslash\hspace{0pt}}m{#1}}

\allowdisplaybreaks

\newcommand{\ev}[1]{\ensuremath{\langle #1 \rangle}}
\newcommand{\abs}[1]{\ensuremath{\lvert#1\rvert}}
\newcommand{\upd}{\ensuremath{\mathrm d}}
\newcommand{\ba}[1]{\ensuremath{\left( #1 \right)}}
\newcommand{\bb}[1]{\ensuremath{\left[ #1 \right]}}
\newcommand{\be}[2]{\ensuremath{\left.  #1 \right|_{#2}}}
\newcommand{\pd}[2]{\ensuremath{\frac{\partial #1}{\partial #2}}}
\newcommand{\pdd}[2]{\ensuremath{\frac{\partial^2 #1}{\partial #2^2}}}
\newcommand{\td}[2]{\ensuremath{\frac{\mathrm d #1}{\mathrm d #2}}} 
\newcommand{\order}[1]{\ensuremath{\mathcal{O}(#1)}}
\newcommand{\nocontentsline}[3]{}
\newcommand{\tocless}[2]{\bgroup\let\addcontentsline=\nocontentsline#1{#2}\egroup}

\newcommand{\hhOmega}[1]{\ensuremath{h^2 \Omega_{#1}}\xspace}
\newcommand{\SNR}{\ensuremath{\rho}\xspace}
\newcommand{\g}{\ensuremath{g_{\star}}\xspace}
\newcommand{\gs}{\ensuremath{g_{\star S}}\xspace}
\newcommand{\gG}{\ensuremath{g_{\star,\text{SM}}}\xspace}
\newcommand{\gsg}{\ensuremath{g_{\star S,\text{SM}}}\xspace}

\newcommand{\DOFnu}{\ensuremath{g_{\nu}}\xspace}
\newcommand{\DOFh}{\ensuremath{g_{h}}\xspace}

\newcommand{\vw}{\ensuremath{v_w}\xspace}
\newcommand{\fp}{\ensuremath{f_p}\xspace}
\newcommand{\Tobs}{\ensuremath{t_\text{obs}}\xspace}
\newcommand{\temprat}[1]{\ensuremath{\xi_{#1}}\xspace}
\newcommand{\tempratinit}{\ensuremath{\xi_h^\text{init}}\xspace}
\newcommand{\Th}{\ensuremath{T_h}\xspace}
\newcommand{\Tg}{\ensuremath{T_\gamma}\xspace}
\newcommand{\Tnudec}{\ensuremath{T_\gamma^\text{$\nu$-dec}}\xspace}
\newcommand{\Thdec}{\ensuremath{T_\gamma^\text{$h$-dec}}\xspace}
\newcommand{\Thann}{\ensuremath{T_\gamma^\text{$h$-ann}}\xspace}
\newcommand{\Theq}{\ensuremath{T_\gamma^\text{$h$-eq}}\xspace}

\newcommand{\Thnuc}{\ensuremath{T_h^\text{nuc}}\xspace}
\newcommand{\Tnuc}{\ensuremath{T_\gamma^\text{nuc}}\xspace}
\newcommand{\Neff}{\ensuremath{N_\text{eff}}\xspace}
\newcommand{\NeffSM}{\ensuremath{N_\text{eff}^\text{SM}}\xspace}
\newcommand{\Nnu}{\ensuremath{N_\nu}\xspace}
\newcommand{\rsparam}{\ensuremath{\mathcal{R}}\xspace}


\begin{document}

\title{Dark, Cold, and Noisy: Constraining Secluded Hidden Sectors\\ with
Gravitational Waves} 

\author{Moritz~Breitbach}\email{breitbach@uni-mainz.de}
\affiliation{PRISMA Cluster of Excellence and
             Mainz Institute for Theoretical Physics,
             Johannes Gutenberg-Universit\"{a}t Mainz, 55099 Mainz, Germany}
\author{Joachim~Kopp}\email{jkopp@cern.ch}
\affiliation{PRISMA Cluster of Excellence and
             Mainz Institute for Theoretical Physics,
             Johannes Gutenberg-Universit\"{a}t Mainz, 55099 Mainz, Germany}
\affiliation{Theoretical Physics Department, CERN, 1211 Geneva, Switzerland}
\author{Eric~Madge}\email{eric.madge@uni-mainz.de}
\author{\\Toby~Opferkuch}\email{opferkuch@uni-mainz.de}
\author{Pedro~Schwaller}\email{pedro.schwaller@uni-mainz.de}
\affiliation{PRISMA Cluster of Excellence and
             Mainz Institute for Theoretical Physics,
             Johannes Gutenberg-Universit\"{a}t Mainz, 55099 Mainz, Germany}

\date{\today}

\preprint{CERN-TH-2018-255, MITP/18-115}

\begin{abstract}
We explore gravitational wave signals arising from first-order phase transitions
occurring in a secluded hidden sector, allowing for the possibility that
the hidden sector may have a different temperature than the Standard Model sector.
We present the sensitivity to such scenarios for both current and future
gravitational wave detectors in a model-independent fashion.  Since secluded hidden
sectors are of particular interest for dark matter models at the MeV scale or below,
we pay special attention to the reach of pulsar timing arrays.
Cosmological constraints on light degrees of freedom restrict the number
of sub-MeV particles in a hidden sector, as well as the hidden sector temperature.
Nevertheless, we find that observable first-order phase transitions can occur.
To illustrate our results, we consider two minimal benchmark models: a model
with two gauge singlet scalars and a model with a spontaneously broken $U(1)$
gauge symmetry in the hidden sector.
\end{abstract}

\maketitle
{
	\hypersetup{hidelinks}
	\tableofcontents
}

\section{Introduction}
\label{sec:intro}

With the advent of gravitational wave (GW) astronomy, we can for the first time
hope to sense physics that happened before the era of Big Bang Nucleosynthesis
(BBN).  In fact, gravitational wave observatories on the ground and in space, as well as pulsar
timing arrays (PTAs), will be sensitive to the reverberations of spacetime
caused by violent first-order phase transitions (PTs) in the very early
Universe.  Though produced almost 14 billion years ago, these perturbations can
still be detectable today, as a contribution to the stochastic gravitational wave
background~\cite{Witten:1984rs,Hogan:1984hx,Hogan:1986qda,Turner:1990rc}. (For
recent work on this topic, see Refs.~\cite{Caprini:2015zlo,Caprini:2018mtu,Mazumdar:2018dfl} and
references therein.)  The gravitational wave amplitude and frequency spectrum allow us to
infer both the properties of the phase transition that generated them and the
history of the Universe since then.

It is this connection between stochastic gravitational waves and the particle physics
responsible for their emission that we focus on in this paper.  In particular,
we study scenarios in which gravitational waves are emitted when a scalar field transitions from a local minimum of its potential into the global minimum. We are especially
interested in the case that the scalar is part of a \enquote{hidden sector},
i.e.\ a group of particles that interact only very weakly with Standard Model
(SM) particles and are therefore extremely challenging to detect directly.
In the extreme case where the hidden sector is coupled only via gravity,
gravitational waves may provide the only way of studying such new physics scenarios.

In this paper, we expand on past studies of gravitational waves from
hidden sector phase transitions~\cite{Schwaller:2015tja, Jaeckel:2016jlh, Addazi:2016fbj,
Baldes:2017rcu, Addazi:2017gpt, Tsumura:2017knk, Aoki:2017aws, Geller:2018mwu,
Croon:2018erz, Baldes:2018emh} in several ways:
\begin{enumerate}
  \item[(i)] {\bf Light hidden sectors.}
    We pay special attention to the possibility that the hidden sector may
    contain interesting dynamics at sub-MeV temperatures.  New particles in the
    mass range $\lesssim \text{MeV}$ are motivated for instance by the lack of
    direct detection signals of dark matter. Moreover, gravitational wave signals
    from low-temperature phase transitions may be easier to detect because the
    ratio of the gravitational wave energy to the total energy density of the Universe
    (given by the parameter $\alpha$ that will be defined and discussed in
    \cref{sec:gw-params}) can be larger.  We will study the interplay
    between cosmological constraints on sub-MeV particles (in particular the
    upper limit on the total relativistic energy density, as measured by the
    parameter $\Neff$) on the one hand, and the observability conditions for
    stochastic gravitational waves on the other hand. This will, in particular,
    require us to investigate in detail the sensitivity of PTAs, which are sensitive to gravitational waves in the relevant frequency
    range $\sim 10^{-9}$--$10^{-7}$\,Hz.  A hidden sector phase
    transition will only be observable if a sufficiently large amount of energy
    is converted into gravitational waves. This sets a lower limit on the hidden
    sector energy density.  We will present our results as limits on the number
    of hidden sector degrees of freedom and on the hidden sector temperature.

  \item[(ii)] {\bf Unequal photon and hidden sector temperatures.}
    In the past, it has usually been assumed that the temperatures of
    the hidden and visible sectors are the same.  This, of course, need not be
    the case: because of their weak coupling, the two sectors may never come
    into thermal contact.  And even if they are heated to the same temperature
    after inflation, their subsequent evolution may be different.  In
    particular, whenever a heavy particle species becomes non-relativistic and
    annihilates or decays into lighter species, the corresponding entropy dump
    into the lighter species delays Hubble cooling.  If the numbers of heavy
    particles are very different in the two sectors, this alone can already
    lead to substantial temperature differences. For phase transitions at low
    (sub-MeV) temperatures detectable in PTAs, a temperature difference of the
    hidden sector with respect to the SM sector is even required, otherwise the 
    additional light degrees of freedom would be inconsistent with cosmology.
    We will therefore
    investigate how such temperature differences can affect gravitational wave
    signals, their detectability, and their interpretation.

  \item[(iii)] {\bf Toy models.}  We construct two explicit toy models that feature
    a hidden sector phase transition at sub-MeV scales. We demonstrate that observable gravitational wave signals are possible for narrow strips of parameter space even in light of stringent constraints on additional light degrees of freedom. The first model
    involves a hidden sector consisting of two scalar fields, the second one
    is a Higgsed dark photon model.  Due to the aforementioned restrictions
    on the number of light degrees of freedom in the hidden sector, these toy
    models should cover a large fraction of the model space at sub-MeV
    energies.  We expect that almost any perturbative hidden sector that has
    been in thermal contact with the SM sector not too far above the electroweak
    scale and that features an observable first-order phase transition at sub-MeV
    temperatures should reduce to one of our toy models, or simple variants
    thereof.
\end{enumerate}

We begin in \cref{sec:sensitivity} with a careful discussion of stochastic
gravitational wave spectra from phase transitions in the early Universe.
We then translate these spectra into
sensitivity curves for current and future gravitational wave detectors.  This
model-independent discussion is complemented in \cref{sec:toymodels} by the
introduction of our two toy models, and by the discussion of future gravitational
wave constraints on these models.  Our conclusions will be summarized in
\cref{sec:conclusions}.

\section{Gravitational Wave Sensitivity}
\label{sec:sensitivity}

\subsection{Hidden Sector Cosmology}
\label{sec:cosmology}

We begin our discussion by reviewing important constraints on hidden sectors,
paying special attention to the possibility that the hidden and visible sectors
have different temperatures. We define the temperature ratio
\begin{align}
    \temprat{h} \equiv \frac{\Th}{\Tg} \,,
\end{align}
where $\Th$ and $\Tg$ are the hidden sector and visible sector (photon)
temperatures, respectively.  A temperature ratio $\temprat{h} \ne 1$ could have
been generated already during reheating after inflation, for instance
if inflatons decay preferentially into one of the two sectors, and the two sectors
never come into thermal contact.  Another possible source of temperature
differences is early decoupling of two sectors that initially share the same
temperature.  Whenever a thermalized particle species becomes non-relativistic in
one sector, the entropy associated with it is transferred to the remaining
particle species, thereby heating that sector up.  If one sector contains
significantly more heavy particles than the other, values of $\temprat{h}$ very
different from unity can be expected.  In what follows, we will be interested in
the case $\temprat{h} < 1$, as we will see that some cosmological
constraints are relaxed if the hidden sector is colder than the photons.

This is in particular true for hidden sectors containing particles at the MeV
scale or below. (Heavier hidden sectors are essentially unconstrained.)
As we will see, sub-MeV hidden sectors are of
particular interest for detection in PTAs.  The main constraints on hidden
particles with masses $\lesssim \text{MeV}$ arise from measurements of the
relative abundance of
light elements from Big Bang Nucleosynthesis (BBN)~\cite{Aver:2015iza,
Peimbert:2016bdg}, as well as measurements of the Cosmic Microwave Background
(CMB) power spectrum~\cite{Ade:2015xua, Aghanim:2018eyx}.  A conflict with
these measurements could arise for two reasons: First, the annihilation of a
hidden sector species during BBN or recombination might heat up the photon or
the neutrino bath, affecting the predicted light element abundances in 
a detectable way. Second, light hidden sector particles (``dark
radiation'') can directly affect the expansion rate of the Universe because the
latter is proportional to the total energy density, which in turn is dominated
by contributions from relativistic particles until just before recombination.
These contributions are parameterized by the \textit{effective number of
neutrino species}, defined as
\begin{align}
  \Neff &\equiv \left( \frac{8}{7} \frac{\rho_R - \rho_\gamma}{\rho_\gamma} \right)
                \bigg( \frac{11}{4} \bigg)^{4/3} \,,
  \label{eq:Neff}
\end{align}
where $\rho_R$ and $\rho_\gamma$ are the energy densities of all relativistic
species and of the photons, respectively.  In the
SM, $\NeffSM = 3.046$~\cite{Mangano:2005cc}.
The most conservative 95\% confidence level (CL) constraint on \Neff from 2018
Planck data is
$\Neff = 3.00 \substack{+0.57\\ -0.53}$ at
recombination, while an analysis taking into account data on CMB polarization and
baryon acoustic oscillations (BAO) yields~\cite{Aghanim:2018eyx} 
\begin{align}
  \Neff &= 2.99 \substack{+0.34\\ -0.33} \,.
        & \text{(Planck TT,\,TE,\,EE\,+\,lowE\,+\,lensing\,+\,BAO)}
  \label{eq:Neff-constraint-CMB}
\end{align}
In view of the current tension between the high and low redshift measurements
of the Hubble parameter $H_0$ (see \cite{0004-637X-855-2-136} and
references therein), this bound may be
considered somewhat too strong.
Taking all measurements of $H_0$ at face value, the bound is relaxed to
\begin{align}
  \Neff &= 3.27 \pm 0.30 \,.
        & \text{(Planck TT,\,TE,\,EE\,+\,lowE\,+\,lensing\,+\,BAO\,+\,$H_0$)}
  \label{eq:Neff-constraint-CMB-H0}
\end{align}
Formation of light nuclei during BBN imposes the complementary 95\%~CL
constraint~\cite{Aghanim:2018eyx}
\begin{align}
  \Neff = 2.95 \substack{+0.56\\ -0.52} \,,
  \label{eq:Neff-constraint-BBN}
\end{align}
assuming \Neff remains constant during BBN
\cite{Shvartsman:1969mm,Steigman:1977kc,Scherrer:1987rr,Patrignani:2016xqp} (see
for instance Ref.~\cite{Hufnagel:2017dgo} for the impact of relaxing this assumption).

It is convenient to express $\rho_R$ in \cref{eq:Neff} in terms of hidden
sector parameters, in particular the temperature \Th and the number of
effective relativistic hidden sector degrees of freedom (DOFs) \DOFh.
``Effective'' here means that each bosonic degree of freedom contributes one
unit to \DOFh, while each fermionic degree of freedom contributes $\frac{7}{8}$.
This way, \DOFh can be used directly in the computation of the relativistic energy
density.  We can distinguish the following cases:
\begin{enumerate}
  \item {\bf Hidden sector in thermal contact with the SM.}
    A hidden sector that is in thermal contact with the photons throughout the
    BBN (and $e^\pm$-annihilation) epoch is disfavored by both the BBN and
    CMB$+H_0$
    constraints on
    $\Neff$: \cref{eq:Neff-constraint-CMB-H0} shows that even a single real scalar
    degree of freedom ($g_h = 1$) would be inconsistent with the constraint at
    95\%~CL. We will therefore not consider this scenario in the following. 

    The BBN constraint on $\Neff$ is somewhat weaker for a hidden sector which
    is in thermal contact with the photons only via the neutrinos at the
    temperature $\Tnudec \sim \SI{}{\MeV}$ of neutrino
    decoupling~\cite{Baumann:2018muz}, i.e.\ which couples efficiently to
    neutrinos, but not directly to photons. In this case, we find
    \begin{align}
      \Neff &= \NeffSM \Big( 1 + \frac{\DOFh}{\DOFnu} \Big)
    \end{align}
    before the hidden sector particles become non-relativistic and annihilate
    or decay away into neutrinos, and
    \begin{align}
      \Neff &= \NeffSM \Big( 1 + \frac{\DOFh}{\DOFnu} \Big)^{4/3}
    \end{align}
    afterwards~\cite{Boehm:2012gr}.  In these expressions $\DOFnu = 2 \cdot
    \Nnu \cdot \frac{7}{8} = \frac{21}{4}$ is the effective number of SM
    neutrino
    degrees of freedom,
    and $\Nnu = 3$ is the number of SM neutrino generations. Here $\DOFh$
    is understood to mean the number of relativistic degrees of freedom in the hidden sector at
    $\Tnudec$.  The CMB$+H_0$ constraint on \Neff given in
    \cref{eq:Neff-constraint-CMB-H0} translates to the bound $\DOFh \lesssim
    0.90$ ($\lesssim 0.66$) at $95\%$~CL if the hidden
    sector becomes non-relativistic after (before) recombination. In other
    words, no new light degree of freedom can remain in thermal equilibrium with the SM
    neutrinos after $\Tnudec$, even if we use the least stringent of the three cosmology constraints under consideration.

  \item {\bf Completely decoupled hidden sector.}
    If the hidden sector starts out at early times with a temperature different
    from the SM bath and never (re\mbox{-})enters thermal equilibrium with the visible
    sector, the additional contribution to \Neff is
    \begin{align}
      \Neff &= \NeffSM  + \frac{4}{7} \bigg( \frac{11}{4} \bigg)^{4/3} \DOFh \,
                          \temprat{h}^4 \,.
      \label{eq:Neff-dec}
    \end{align}
    where \DOFh is now understood to be the effective number of relativistic
    hidden sector degrees of freedom at hidden sector temperature $\Th=
    \temprat{h}T_\gamma$.  The same expression holds if the hidden sector was
    in equilibrium with the SM sector at early times, but later
    decoupled while still relativistic. 
        
    In the left-hand panel of \cref{fig:Neff-constraints} we show the CMB and
    BBN constraints on the hidden sector parameters $\temprat{h}$ and $\DOFh$ based on \cref{eq:Neff-dec}.  The orange and blue regions are
    excluded at $2\sigma$ by the BBN and CMB$+H_0$ constraints, respectively,
    while the CMB only bound is shown as a dashed-blue line.  Wee see that hidden
    sectors with more degrees of freedom need to be colder to keep the total
    hidden sector energy density (and thus $\Neff$) consistent with the constraints.
    Even a hidden sector with only one additional degree of freedom ($\DOFh =
    1$) is excluded unless the hidden sector temperature is smaller than $0.6$
    times the SM temperature.

    Note that for a completely decoupled hidden sector, an important additional
    constraint may arise from the total matter density of the Universe. As hidden sector
    particles become non-relativistic, care must be taken that their energy density
    gets converted into a form of dark radiation in order not to overclose the Universe.

  \begin{figure}
    \includegraphics[width=0.495\textwidth]{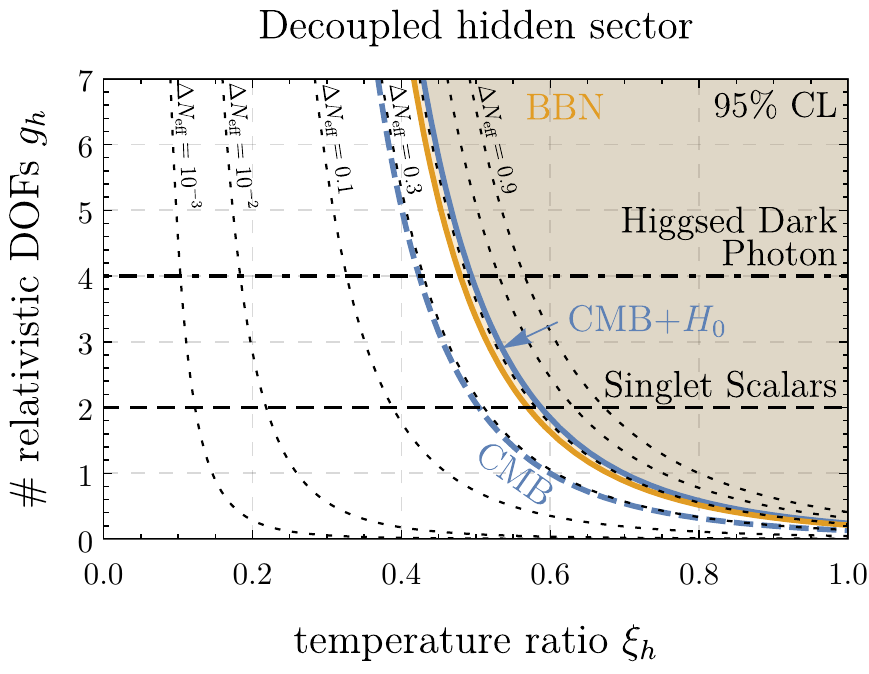}
    \includegraphics[width=0.495\textwidth]{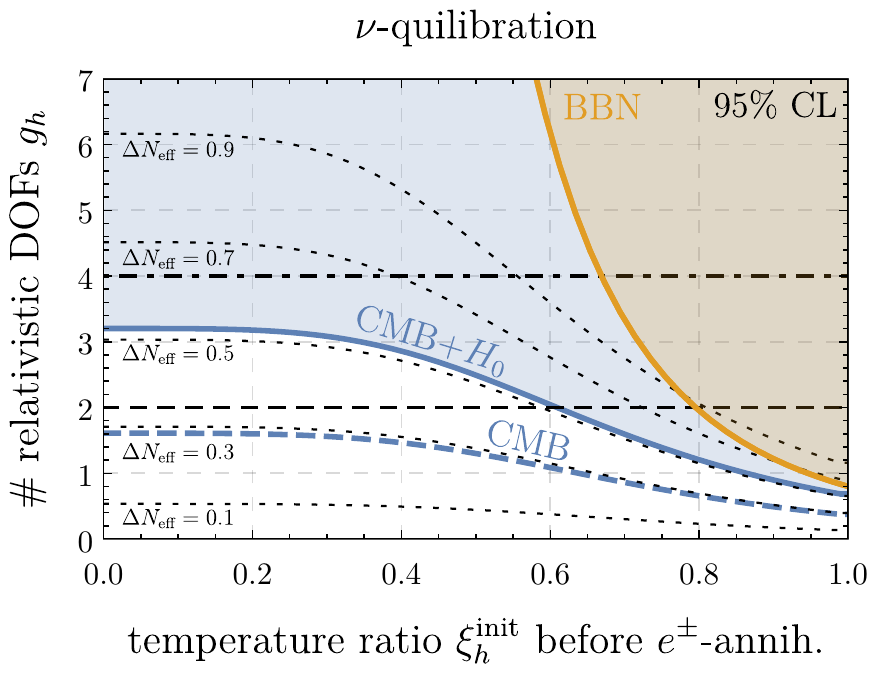}
    \caption{Black dotted contours: the change in $\Neff$ as a function of the effective
      number of relativistic degrees of
      freedom in the hidden sector ($\DOFh$) and of the ratio between the
      hidden and visible sector temperatures ($\temprat{h}$).
      The left-hand panel is for the case of a hidden sector that is never
      in thermal equilibrium with SM particles (or has decoupled long before
      BBN).  The right-hand panel is for the \enquote{$\nu$\mbox{-}quilibration} scenario, where
      the hidden sector (re\mbox{-})couples with the neutrinos, after the latter have
      decoupled from the photons.  The horizontal axis in the right-hand panel
      gives $\tempratinit$, the hidden sector-to-photon temperature ratio
      before both (re\mbox{-})coupling and $e^\pm$\mbox{-}annihilation. The orange shaded
      region indicates the BBN constraint on $\Delta\Neff$ from
      \cref{eq:Neff-constraint-BBN}, while the blue shaded region and blue
      dashed contour show the CMB constraints with and without the low-redshift
      measurements of $H_0$, see
      \cref{eq:Neff-constraint-CMB,eq:Neff-constraint-CMB-H0}. The horizontal
      dashed and dot-dashed lines indicate the number of degrees of freedom in
      the two toy models discussed in \cref{sec:toymodels}. These figures serve
      as an update to Refs.~\cite{Berlin:2017ftj, Berlin:2018ztp} based on the
      new Planck results~\cite{Aghanim:2018eyx}.}
    \label{fig:Neff-constraints}
  \end{figure}

  \item {\bf Hidden sector equilibrates with neutrinos ($\nu$\mbox{-}quilibration)}.
    While the hidden sector cannot be in thermal contact with the photon bath
    at temperatures around or below the BBN temperature $T_\text{BBN}$, there
    is the possibility that it (re\mbox{-})couples with neutrinos after the latter have
    lost thermal contact with the photons \cite{Berlin:2017ftj, Berlin:2018ztp}. This scenario, which we will dub
    ``$\nu$\mbox{-}quilibration'', corresponds to the following sequence of events:
    \begin{center}
      \includegraphics{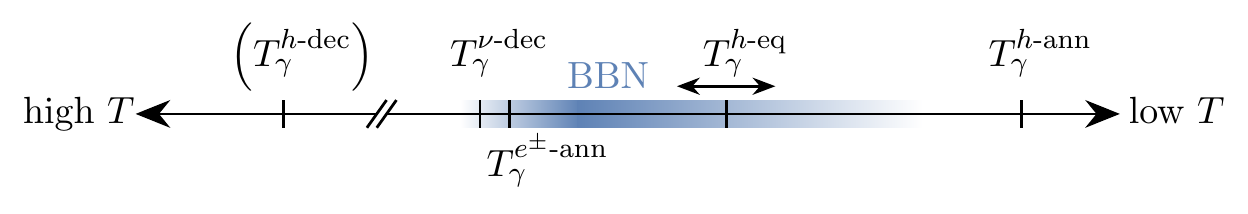}
    \end{center}
    where $\Tnudec \sim \si{\MeV}$ is the photon temperature
    at neutrino decoupling,
    $\Theq$ is the (photon) temperature at which the hidden sector comes into
    thermal contact with neutrinos, and $\Thann$ is the (photon) temperature at
    which the hidden sector becomes non-relativistic and its entropy gets dumped
    back into the neutrino bath. If the hidden sector was ever in thermal
    contact
    with the SM at very early times (long before BBN), we denote the temperature
    at which it lost thermal contact by $\Thdec$.  Assuming that all hidden
    sector particles are relativistic when the hidden sector comes into thermal
    equilibrium with the neutrinos, and that equilibration is
    quasi-instantaneous, the temperature change experienced by the
    equilibrating sectors is determined by conservation of energy density. If
    the temperature ratio between the hidden sector and the photons is
    $\tempratinit$ prior to BBN, $e^\pm$\mbox{-}annihilation, and hidden
    sector--neutrino (re\mbox{-})coupling, the temperature ratio after (re\mbox{-})coupling
    is
    \begin{align}
      \temprat{\nu+h} &= \temprat{\nu}^\text{SM}
                         \bigg[1 + \frac{\DOFh}{\DOFnu} \bigg]^{-\frac{1}{4}}
                         \bigg[
                           1 + \frac{\DOFh}{\DOFnu} (\tempratinit)^4
                         \bigg]^{\frac{1}{4}} ,
                      &  \left(\begin{array}{l}
                           \text{after hidden sector} \\
                           \text{thermalization with $\nu$'s}
                         \end{array} \right) ,
      \label{eq:xi-reeq-1}
    \intertext{%
      where $\temprat{\nu}^\text{SM}$ is the temperature ratio between the
      neutrino and photon sectors in the Standard Model. Its value is
      approximately, but not exactly, $(4/11)^{1/3}$. A small correction factor
      $(\NeffSM / \Nnu)^{1/4}$ arises because of the effect of $e^\pm$
      \mbox{-}annihilation on the neutrino temperature. \Cref{eq:xi-reeq-1} is valid
      assuming that $e^\pm$\mbox{-}annihilation finishes before the hidden sector
      equilibrates.  The effective number of neutrino species is modified to}
      \Neff &= \NeffSM \bigg[ 1 + \frac{\DOFh}{\DOFnu} (\tempratinit)^4 \bigg] \,,
            &  \left( \begin{array}{l}
                        \text{after hidden sector} \\
                        \text{thermalization with $\nu$'s}
                      \end{array} \right) .
      \label{eq:Neff-reeq-1}
    \end{align}
    This shows that \Neff returns to \NeffSM in the limit
    $\tempratinit \to 0$.

    When the neutrino/hidden sector temperature drops further below $\Theq$,
    there are two possibilities for the decoupling of the hidden sector from the
    neutrinos. The first possibility is that interactions between the neutrinos
    and the hidden sector particles become too small to maintain equilibrium,
    resulting in decoupling while the hidden sector is still relativistic. As
    the hidden sector subsequently becomes non-relativistic, entropy cannot be
    transferred to the neutrino bath, so the neutrino-to-photon temperature
    ratio stays at the value given by \cref{eq:xi-reeq-1}.
    \Neff at recombination, however, will have decreased
    compared to \cref{eq:Neff-reeq-1} by a factor $\DOFnu / (\DOFnu + \DOFh)$,
    possibly violating CMB constraints. Even more problematic, the frozen out
    non-relativistic hidden sector particles will contribute to the dark matter
    density, most likely overclosing the Universe. Therefore, we will not consider
    this case any further.

    The second possibility is that the neutrino/hidden sector temperature drops
    below the masses of
    the hidden sector particles while the two sectors are still in equilibrium.
    The hidden sector particles will then annihilate back to neutrinos, thereby
    reheating the neutrino bath with respect to the photons. Beginning with 
    \cref{eq:xi-reeq-1,eq:Neff-reeq-1} and using conservation of
    co-moving entropy yields
    \begin{align}
      \temprat{\nu} &= \temprat{\nu}^\text{SM}
                       \bigg[1 + \frac{\DOFh}{\DOFnu} \bigg]^{\frac{1}{12}}
                       \bigg[
                         1 + \frac{\DOFh}{\DOFnu} (\tempratinit)^4
                       \bigg]^{\frac{1}{4}} \,,
                    &  \left(\begin{array}{l}
                         \text{after hidden sector} \\
                         \text{annihilation into $\nu$'s}
                       \end{array} \right),
      \label{eq:xi-reeq-2}
    \intertext{for the neutrino-to-photon temperature ratio, and}
      \Neff &= \NeffSM \bigg[ 1 + \frac{\DOFh}{\DOFnu} \bigg]^{\frac{1}{3}}
                       \bigg[ 1 + \frac{\DOFh}{\DOFnu} (\tempratinit)^4
                       \bigg] \,,
            &  \left( \begin{array}{l}
                        \text{after hidden sector} \\
                        \text{annihilation into $\nu$'s}
                      \end{array} \right),
      \label{eq:Neff-reeq-2}
    \end{align}
    for the effective number of neutrino species after hidden sector
    decoupling. Unless hidden sector particles have masses $\lesssim 
    \text{eV}$, this is the value that will be measured in the CMB.
    
    In the right-hand panel of \cref{fig:Neff-constraints} we show the excluded
    regions in the $\tempratinit$--$\DOFh$ plane for the $\nu$\mbox{-}quilibration
    scenario.  We remind the reader that $\tempratinit$, plotted on the
    horizontal axis, is the hidden sector-to-photon temperature ratio before
    both BBN and $e^\pm$\mbox{-}annihilation. As before, the orange and blue regions
    are excluded at 95\%~CL by BBN and CMB$+H_0$ constraints, respectively,
    while the CMB-only constraint is given by the dashed blue line. BBN
    constraints are based on \cref{eq:Neff-reeq-1}, which corresponds to the
    assumption that hidden sector particles become non-relativistic only after
    BBN, or, more precisely, after the initial formation of the light elements
    (${}^4$He and D) at $\Tg \sim 0.1$~MeV \cite{Cyburt:2015mya}.  The CMB
    constraint is derived from \cref{eq:Neff-reeq-2}.
\end{enumerate}

\subsection{Temperature Dependence of the Gravitational Wave Parameters}
\label{sec:gw-params}

We have concluded from \cref{fig:Neff-constraints} that hidden sectors with a
large particle content at masses $\lesssim \text{MeV}$ may be
cosmologically allowed, but only if their temperature is substantially lower
than the photon temperature ($\temprat{h} < 1$), and their coupling to photons
is tiny. The second of these requirements makes it extremely hard to probe such
hidden sectors using conventional cosmological, astrophysical, or laboratory
methods.  This is why we focus here on gravitational waves: if the hidden
sector undergoes a strongly first-order phase transition at some point in
cosmological history, the gravitational waves emitted in the process may be
detectable today as part of the stochastic gravitational wave background. The
frequency dependence of the gravitational wave spectrum furthermore contains
information about the properties of the phase
transition and the subsequent expansion history of the Universe.

The spectra of gravitational waves sourced by cosmological first-order
phase transitions depend
mainly on three parameters: the strength of the phase transition, $\alpha$,
its inverse time scale, $\beta$, and the temperature $T^\text{nuc}$ at
which the transition occurs (nucleation temperature)~\cite{Turner:1992tz,
Kamionkowski:1993fg, Grojean:2006bp, Caprini:2015zlo}.  In the following,
we will define these parameters in terms of particle physics quantities.
For phase transitions occurring in a hidden sector, we will in particular
discuss their dependence on the temperature ratio $\temprat{h}$.

A first-order phase transition proceeds by the nucleation of bubbles of the new
(true vacuum) phase in the background plasma that is still in the old
(false vacuum) phase. Thus, the transition begins when the rate $\Gamma$ at
which such bubbles form within a Hubble volume exceeds the Hubble time. The
temperature at which this happens is the \emph{bubble nucleation temperature}
$T^\text{nuc}$.  The bubble nucleation rate per unit volume is given by $\Gamma(T) =
A(T) e^{-S_E(T)}$~\cite{Coleman:1977py,Callan:1977pt,Linde:1981zj} where $S_E
(T)$ is the temperature dependent Euclidean action corresponding to the
transition from the false vacuum to the true one.  For bubble nucleation at
non-zero temperature the time component of the 4-dimensional Euclidean action
becomes $T^{-1}$, i.e. $S_E(T)\equiv S_3(T)/T$ where $S_3 (T)$ is the remaining
3-dimensional Euclidean action \cite{Linde:1981zj}.  Assuming $A(T) \sim T^4$
and that the phase transition occurs during a radiation dominated epoch such
that $H^2 = \frac{8 \pi G_N}{3} \rho_R$ ($\rho_R$ is the radiation energy
density and $G_N$ is Newton's constant),
the nucleation criterion $\Gamma \sim H^4$ evaluates to 
\begin{align}
  \frac{S_3(T^\text{nuc})}{T^\text{nuc}} \sim
    146 - 2 \log \bigg( \frac{\g(T^\text{nuc})}{100} \bigg)
        - 4 \log \bigg( \frac{T^\text{nuc}}{\SI{100}{GeV}} \bigg) .
  \label{eq:nuc-cond}
\end{align}
Note that here we still assume that all sectors are in thermal equilibrium, and
$\g(T^\text{nuc})$ is the total number of
relativistic degrees of freedom.\footnote{In our analysis, we will use 
\cref{eq:nuc-cond} even for evaluating the
nucleation criterion in scenarios with $\temprat{h} \neq 1$. The
resulting $T^\text{nuc}$ is then rescaled to account for the true value
of $\temprat{h}$. By doing so, we neglect logarithmic corrections to
\cref{eq:nuc-cond} arising from the difference between $\Thnuc$ and
$\Tnuc$.}

The available energy budget for gravitational wave emission is given by the
latent heat $\epsilon$, i.e.\ the absolute change in the energy density, see
Ref.~\cite{Espinosa:2010hh} for additional details. The latent heat release
normalized to the total radiation density of the Universe, $\rho_R$, at the
time of the phase transition gives the {\it strength of the phase transition},
\footnote{Strictly speaking, we should here distinguish between the temperature
  at which gravitational waves are emitted (usually referred to as $T_*$) and
  $T^\text{nuc}$.  The two
  temperatures may be different if the latent heat released during the phase
  transitions heats the plasma considerably compared to its initial
  temperature, as could be the case for instance for a phase transition
  happening during a vacuum-dominated epoch~  \cite{Brdar:2018num,Ellis:2018mja,Prokopec:2018tnq,vonHarling:2017yew,Bruggisser:2018mrt,Konstandin:2011dr,Kobakhidze:2017mru,
  Iso:2017uuu, Bai:2018vik, Megias:2018sxv, Hambye:2018qjv, Caprini:2015zlo,Randall:2006py}.
  However, in the following we will only consider phase transitions occurring
  during radiation domination, and we will therefore set $T_* = T^\text{nuc}$ in the
  rest of this paper.}
\begin{align}
  \alpha \equiv \frac{\epsilon}{\rho_R}
         = \frac{1}{\rho_R}
           \bigg( -\Delta V + T^\text{nuc} \frac{\partial \Delta V}{\partial T}
                                                           \bigg|_{T^\text{nuc}} \bigg) \,.
  \label{eq:alpha}
\end{align}
Here, $\Delta V < 0$
is the change in the potential between the false and the
true vacua. Once again, $T^\text{nuc}$ is understood to mean the temperature
at the time of the phase transition of the sector in which the phase transition occurs.
For a phase transition occurring in the SM (hidden sector), $T^\text{nuc}$
should thus be replaced by the photon (hidden sector) temperature $\Tnuc$ ($\Thnuc$)
at the time of bubble nucleation.

It is intuitively clear that larger $\alpha$ implies larger gravitational wave
amplitudes. If all particle species abundant in the Universe share the same
temperature, $\rho_R$ is simply given by $\rho_R = \pi^2 [\g (\Tnuc)]^4/30$,
where $\g$ is the total number of effective relativistic degrees of freedom. We will instead
be interested in a phase transition occurring in a hidden sector whose
temperature is
different from the photon temperature, in which case the appropriate expression
is instead
\begin{align}
  \begin{split}
    \rho_R &= \frac{\pi^2}{30} (\gG + \DOFh \temprat{h}^4) (\Tnuc)^4  \\
           &= \frac{\pi^2}{30} \Big(\frac{\gG}{\temprat{h}^4} + \DOFh \Big) (\Thnuc)^4 \,,
  \end{split}
  \label{eq:rho-R}
\end{align}
where $\gG$ and $\DOFh$ are the numbers of relativistic degrees of freedom in the
SM sector and the hidden sector, respectively.
As we have seen in \cref{sec:cosmology}, CMB and BBN constraints on \Neff can
only be satisfied if $\temprat{h} \ll 1$. Then, for fixed $\Thnuc$,
$\rho_R$ scales approximately proportional to $\temprat{h}^{-4}$, and consequently
\begin{align}
  \alpha \propto \temprat{h}^4 \,.
\end{align}
In other words, a large temperature ratio between
the visible and hidden sectors
at the time of the phase transition reduces the strength of the gravitational
wave signal compared to the case where both sectors have similar temperatures.

Gravitational wave signals from phase transitions also depend on the
\emph{inverse time scale of the phase transition}, $\beta$: a fast transition
means that many bubbles form simultaneously, so they are still small when they
collide. This results in gravitational wave signals that are weaker and peak at higher
frequencies than
those from a slow transition. The inverse time scale
is given by
\begin{align}
  \beta \equiv - \frac{\upd S_E(t)}{\upd t} \bigg|_{t^\text{nuc}} \,,
\end{align}
where the right-hand side is to be evaluated at the time of bubble nucleation
$t^\text{nuc}$.
It is conventional to normalize $\beta$ to the Hubble rate $H$ at the time of
bubble nucleation, which leads to
\begin{align}
  \frac{\beta}{H} =
    \Thnuc \frac{\upd S_E(T)}{\upd T} \bigg|_{\Thnuc} \,.
  \label{eq:beta}
\end{align}
Note that, unlike $\beta$ itself, $\beta/H$ is independent of the
temperature ratio $\temprat{h}$.

Besides $\Thnuc$, $\alpha$, and $\beta$, gravitational wave spectra in
principle depend also on the \emph{bubble wall velocity}, \vw, i.e.\ the speed
at which bubbles of the true phase expand. We assume an optimistic value $\vw
\sim 1$, which is well justified for strong phase
transitions~\cite{Bodeker:2017cim}. Note that the amplitude of the
gravitational wave power spectrum depends linearly on \vw (see
\cref{sec:gw-spectrum} for the full dependence), therefore small deviations
from $\vw \sim 1$ will not significantly affect our results.

The key result of the above discussion is that a strong gravitational wave
signal and safety from \Neff constraints impose opposite conditions on hidden
sector models: for compatibility with \Neff constraints, $\temprat{h} < 1$ is
needed, while the scaling of $\alpha$ with $\temprat{h}^4$ implies that strong
gravitational wave emission from a hidden sector phase transition is only
possible if $\temprat{h}$ is not too small.  In \cref{sec:toymodels} below, we
will see explicitly the impact of these conflicting conditions on the
parameter space of two toy models. In particular, we will see that for a hidden
sector that is completely decoupled from the SM, observable gravitational wave signals are possible for only thin strips of parameter space. For a hidden sector that (re\mbox{-})enters
thermal equilibrium with the neutrinos after the latter have decoupled,
\cref{eq:xi-reeq-1} shows that the value of $\temprat{h}$ at the phase
transition (after hidden sector (re\mbox{-})coupling) can be significantly larger than
the initial temperature ratio $\tempratinit$. In other words, even though the
initial $\tempratinit$ may be small enough to avoid the BBN constraint on
$\Neff$, the $\temprat{h}$ relevant during the phase transition may be large
enough to yield a strong gravitational wave signal.  Of course, one still needs
to worry about the late-time neutrino-to-photon temperature ratio given by \cref{eq:xi-reeq-2}
possibly violating the CMB constraint on \Neff.

\subsection{Gravitational Wave Spectrum from a Hidden Sector Phase Transition}
\label{sec:gw-spectrum}

In what follows we review the possible mechanisms that source gravitational
waves during a first-order phase transition. We closely follow
Ref.~\cite{Caprini:2015zlo}, but we extend the results from that paper for the
case of a hidden sector temperature differing from the photon temperature. The
production of gravitational waves in a first-order phase transition can be
separated into three stages:
\begin{enumerate}
  \item {\bf Collisions of the bubble walls.} \cite{Huber:2008hg}
    This contribution depends only on the dynamics of the scalar field
    (not on the background plasma) and is therefore often referred to as the
    \emph{scalar field contribution}.
    It is usually treated in the ``envelope approximation'',
    which assumes sizable interactions only at the intersection points of the
    bubble walls and
    a quick dispersion after the bubble collisions.

  \item {\bf Collision of sound waves} in the plasma generated during
  bubble
    expansion~\cite{Hindmarsh:2015qta}. Compared to the contribution arising
    from bubble wall collisions,
    this effect lasts longer and is therefore enhanced by a factor of
    $\beta / H$.\footnote{Note that it currently remains unclear whether the
    simulation of this contribution can be extrapolated to very strong
    transitions with $\alpha > 0.1$~\cite{Hindmarsh:2015qta, Caprini:2015zlo}. In addition, for phase transitions with $\beta/H \gtrsim 100$ (corresponding to a large number of smaller bubbles being nucleated), the timescale of the transition from sound waves to turbulence is expected to be significantly shorter than a Hubble time. Subsequently, the period of time where sound waves can source gravitational waves is cut short. However, current simulations are unable to make precise predictions about the transition to turbulence in the regime $\beta/H \gtrsim 100$. Therefore the exact efficiency of converting sound waves to turbulence remains unknown \cite{Hindmarsh:2017gnf,Ellis:2018mja}.}

  \item {\bf Turbulence in the plasma} after the collision of sounds waves,
    which can last for several Hubble times~\cite{Caprini:2009yp}.
    This contribution is suppressed by a factor of $\varepsilon_\text{turb} =
    5 \sim 10\%$ compared to the sound wave contribution, i.e.\ only a small
    part of the sound wave energy budget is converted into turbulent
    motion~\cite{Hindmarsh:2015qta}.  We will optimistically assume
    $\varepsilon_\text{turb} = 10\%$.
\end{enumerate}
While a complete description of these contributions requires numerical
simulations, it is for our purposes useful to work with an analytic
parameterization for the gravitational wave frequency spectra. A suitable
functional form for the gravitational wave power spectrum at emission is
\cite{Huber:2008hg, Hindmarsh:2015qta, Caprini:2009yp}
\begin{align}\label{eq:spectrum}
  \Omega_\text{GW}(f) \equiv
  \frac{1}{\rho_c} \td{\rho_\text{GW}(f)}{\log f} \simeq
  \mathcal N \, \Delta \,\ba{\frac{\kappa \, \alpha}{1+\alpha}}^p
    \ba{\frac{H}{\beta}}^q  s(f) \,,
\end{align}
where $\rho_c = 3 H^2 / (8 \pi G_N)$ is the critical energy density, with the
Hubble rate in the radiation-dominated era is given by $H^2 = \frac{8 \pi
G_N}{3} \rho_R$.  $\Omega_\text{GW}(f)$ depends on the normalization
factor~$\mathcal N$, the velocity factor~$\Delta$, the efficiency
factors~$\kappa$,
the exponents~$p$ and $q$, and the spectral shape function~$s(f)$. These
parameters are determined for the three contributions separately by fitting to
the results of numerical simulations.  A summary of the resulting parameter
values is given in \cref{tab:gw-spectra}.

\begin{table}
  \centering
  \renewcommand{\arraystretch}{1.3}
  \begin{tabular}{lC{4.5cm}C{4.5cm}C{4.5cm}}
    \toprule
                        & Scalar field $\Omega_\phi$ & Sound waves $\Omega_\text{sw}$ & Turbulence $\Omega_\text{turb}$           \\
    \hline
    $\mathcal N$
                        & $\num{1}$                  & $\num{1.59e-1}$                & $\num{2.01e1}$                            \\
    $\kappa$            & $\kappa_\phi$              & $\kappa_\text{sw}$             & $\varepsilon_\text{turb} \kappa_\text{sw}$ \\
    $p$                 & $2$                        & $2$                            & $\frac{3}{2}$                             \\
    $q$                 & $2$                        & $1$                            & $1$                                       \\
    $\Delta$            & $\frac{0.11\vw^3}{0.42+\vw^2}$ & $\vw$                      & $\vw$                                     \\
    $\fp$       & $\frac{0.62\beta}{1.8-0.1\vw+\vw^2}$
                                                     & $\frac{2\beta}{\sqrt{3}\vw}$   & $\frac{3.5\beta}{2\vw}$                   \\
    $s(f)$              & $\frac{3.8(f/\fp)^{2.8}}{1+2.8(f/\fp)^{3.8}}$ 
                                                     & $(f/\fp)^{3}\ba{\frac{7}{4+3(f/\fp)^{2}}}^{7/2}$
                                                           & $\frac{(f/\fp)^3}{
                                                           (1+f/\fp)^{11/3} [1 + 8\pi (f/H)]}$ \\ \hline
    Reference           & \cite{Huber:2008hg}        & \cite{Hindmarsh:2015qta}       & \cite{Caprini:2009yp} \\
    \botrule
  \end{tabular}
  \caption{Parameters of the gravitational wave spectra, using the phenomenological
    parameterization from \cref{eq:spectrum}.  In addition to the normalization
    factor~$\mathcal N$, the velocity factor~$\Delta$, the efficiency
    factor~$\kappa$ (see discussion around \cref{eq:alpha-inf,eq:kappa-alpha}),
    the exponents~$p$ and $q$, and the spectral shape function~$s(f)$, we also
    list the resulting peak frequency \fp.  In the last three rows, $\vw$
    denotes the bubble wall velocity and $\beta$ is the inverse time scale of
    the transition (see \cref{eq:beta}).  Note that $\mathcal N$
    and $f_p$ listed here are the amplitude and peak frequency at the
    time of the phase transition (hidden sector temperature \Thnuc) and need to
    be redshifted using \cref{eq:full-redshift} to obtain the corresponding values today.}
  \label{tab:gw-spectra}
\end{table}

After its emission, the stochastic gravitational wave background propagates
freely, undisturbed until today.
The expansion of the Universe redshifts both the energy density and
frequency such that the gravitational wave power spectrum today, $\Omega_
\text{GW}^0(f)$, as a function of today's frequency $f$ is
\begin{align}
  \label{eq:full-redshift}
  \Omega_\text{GW}^0(f) &= \mathcal{R}\,\Omega_\text{GW}\left(\frac{a_0}{a}f\right)
\end{align}
where
\begin{align}
  \rsparam &\equiv \bigg(\frac{a}{a_0} \bigg)^4
       \bigg(\frac{H}{H_0} \bigg)^2                     = \bigg(\frac{\gs^\text{EQ}}{\gs} \bigg)^{4/3}
       \bigg(\frac{\Tg^0}{\Tg} \bigg)^4
       \bigg(\frac{H}{H_0}\bigg)^2                      = \bigg(\frac{\gs^\text{EQ}}{\gs} \bigg)^{4/3}
       \frac{8 \pi G_N}{3} \,
       \frac{\g \pi^2 (\Tg^0)^4}{30 H_0^2}              \nonumber\\
    &  \simeq \num{2.473E-5}{} h^{-2}
       \bigg( \frac{\gs^\text{EQ}}{\gs} \bigg)^{4/3} \left(\frac{\g}
       {2}\right)\,.
  \label{eq:amp-redshift}
\end{align}
In the above expressions, $a$ is the scale factor of the Universe at the time
of the phase transition, and $a_0$ is the scale factor today. Similarly $H$ and
$H_0$ ($\Tg$ and $\Tg^0 \simeq
\SI{2.35e-13}{GeV}$~\cite{Fixsen:2009ug,Patrignani:2016xqp}) denote the Hubble
rates (photon temperatures) at the time of the phase transition and today,
respectively. As usual, we denote Newton's constant as $G_N$, and we abbreviate
$H_0 / (\SI{100}{\km~\mega\pc^{-1}\s^{-1}})$ as $h$.  The quantities $\g$
($\gs$) denote the effective numbers of degrees of freedom relevant for the
computation of energy densities (entropy densities).  Without a superscript,
they are meant to be evaluated at the time of the phase transition, while
$\gs^\text{EQ}$ should be evaluated at the time of matter--radiation equality.
(From matter--radiation equality until today the number of degrees of freedom
in the photon bath does not change.) These quantities receive contributions
from the SM sector ($\gG$, $\gsg$) and from the hidden sector according to
\begin{align}
  \g            &= \gG + \DOFh \temprat{h}^4 \,,      
                                        \label{eq:g-star} \\[0.1cm]
  \gs           &= \gsg + \DOFh \temprat{h}^3 \,,    
                                        \label{eq:gS-star} \\
  \gs^\text{EQ} &= 2 + \frac{7}{4} \Nnu (\temprat{\nu}^\text{EQ})^3
                     + g_h (\temprat{h}^\text{EQ})^3 \,.
                                        \label{eq:gS-star-EQ}
\end{align}
In the last expression, $\temprat{\nu}^\text{EQ}$ and $\temprat{h}^\text{EQ}$
refer to the neutrino-to-photon and hidden sector-to-photon temperature ratios
at matter--radiation equality, respectively.  $\temprat{\nu}^\text{EQ}$ may
differ from its SM value if some or all of the hidden sector energy density has
been dumped into the neutrino sector, as for instance in the $\nu$\mbox{-}quilibration
scenario introduced in \cref{sec:cosmology}.
The value of $\temprat{h}^\text{EQ}$ is model-dependent if the
hidden sector is fully decoupled; in this case, it depends on the way in which
the energy density carried by massive ($\gg \text{eV}$) particles gets
transferred to relativistic species.  In the subsequent sections and all
figures we drop the superscript `0', i.e.\ $\Omega_\text{GW}(f)$ will always
refer to the observable spectrum today.

The efficiency factors $\kappa$ of the three contributions to the gravitational
wave spectrum are functions of $\alpha$, and they further depend on the
coupling between the plasma and the bubble wall.  Stronger coupling means that
energy is transferred more efficiently from the expanding bubble wall
into the plasma, increasing the gravitational wave energy radiated by sound waves and
turbulence, and decreasing the energy radiated in bubble collisions.
$\kappa$ is conveniently expressed in terms of a critical phase transition
strength $\alpha_\infty$ that separates the runaway regime ($\alpha >
\alpha_\infty$) in which bubble walls are accelerated continuously, and the
non-runaway regime ($\alpha < \alpha_\infty$) in which they reach a terminal
velocity. Note, however, that it was argued in Ref.~\cite{Bodeker:2017cim} that gauge
bosons gaining mass in the phase transition prevent bubble walls from reaching
the runaway regime, even if $\alpha > \alpha_\infty$.  Note also that bubble
wall velocities $\vw \sim 1$ can and will still be reached even in the non-runaway
regime for sufficiently strong transitions~\cite{Bodeker:2017cim}.  For a
phase transition in a hidden sector, $\alpha_\infty$ is given
by~\cite{Espinosa:2010hh, Caprini:2015zlo}
\begin{align}
  \alpha_\infty
    &\equiv \frac{(\Thnuc)^2}{\rho_R(\Tnuc)}
              \bigg[ \sum_{i=\text{bosons}}   n_i \frac{\Delta m_i^2}{24} +
                     \sum_{i=\text{fermions}} n_i\frac{\Delta m_i^2}{48} \bigg] \,,
  \label{eq:alpha-inf}
\end{align}
where the sums run over hidden sector bosons and fermions which gain mass during the phase transition, $n_i$
is the physical number of degrees of freedom of the $i$-th hidden sector
particle, and $\Delta m_i>0$ is the change in its mass.  In the non-runaway regime,
the scalar field contribution to the gravitational wave spectrum is negligible
($\kappa_\phi = 0$) as the latent heat released in the phase transition is
efficiently converted into plasma motion.  The efficiency factor for the sound
wave contribution is then given by~\cite{Espinosa:2010hh}
\begin{align}
  \kappa_\text{sw} = \kappa(\alpha)
                  \simeq \frac{\alpha}{0.73 + 0.083 \sqrt{\alpha} + \alpha}
  \label{eq:kappa-alpha}
\end{align}
in the case $\vw\sim 1$. In the runaway regime, a fraction $\alpha_\infty / \alpha$ of the latent heat
is converted into plasma motion, and the remainder goes into further
accelerating the bubble wall.  Thus, $\kappa_\text{sw} = \kappa(\alpha_\infty)
\, \alpha_\infty / \alpha$ and $\kappa_\phi = 1 - \alpha_\infty/\alpha$ in the
runaway regime.  A fraction $\varepsilon_\text{turb}$ of the bulk motion energy
is finally converted into turbulent motion, i.e.\ $\kappa_\text{turb} =
\varepsilon_\text{turb} \kappa_\text{sw}$, where we take
$\varepsilon_\text{turb} = 10\%$.

The numerical simulations from which these parameter values are obtained are
based on the assumption that the phase transition is happening in the SM
sector, which, for the temperature ranges under consideration, always contains
at least one relativistic degree of freedom, ensuring a speed of sound close to
the speed of light.  This is not necessarily the case for a decoupled hidden
sector, and our results will \emph{not} apply to hidden sectors that have only
massive, non-relativistic degrees of freedom immediately after the phase
transition.

\subsection{Experimental Noise Curves and Power-Law Integrated Sensitivities}
\label{sec:sensitivities}

\begin{figure}
  \makebox[\textwidth][c]{\includegraphics[width=1.125\textwidth, trim={0cm
  0.4cm 0cm 0.38cm}, clip]{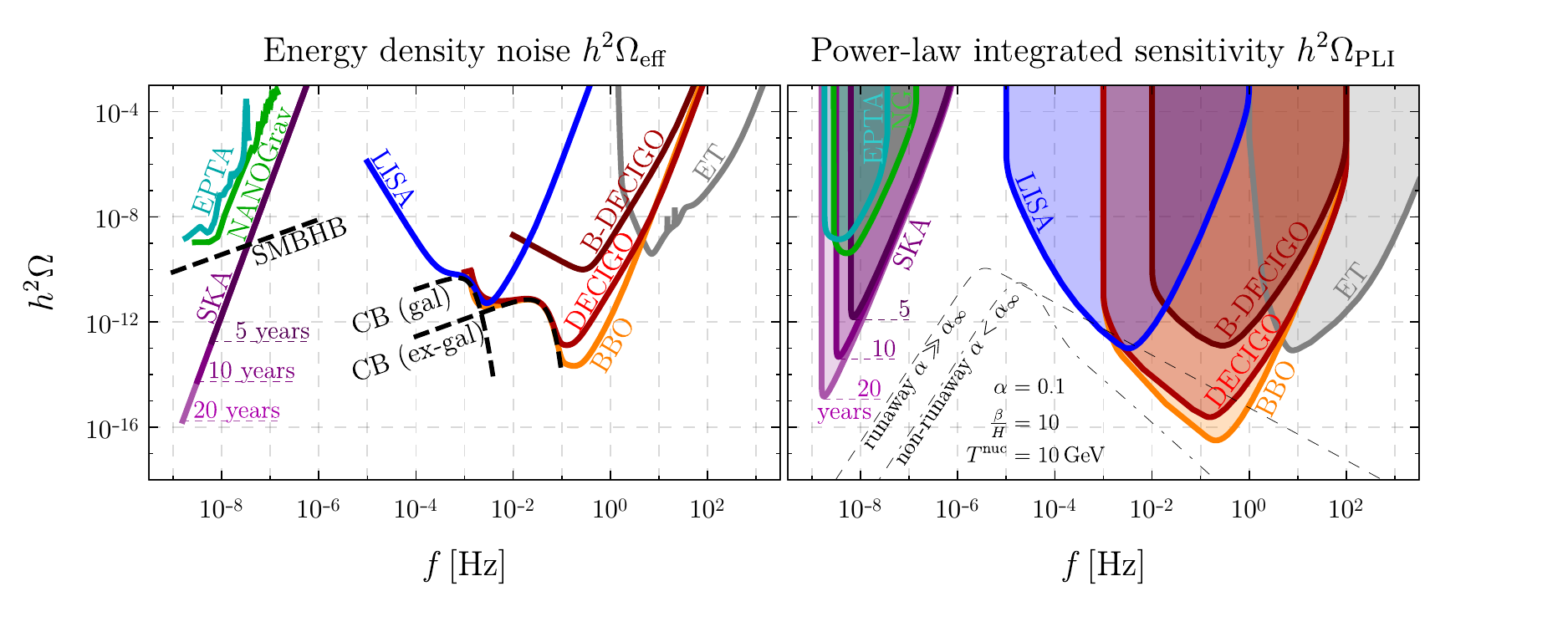}}
  \caption{Noise curves (left) and PLI sensitivity curves
    (right) for various gravitational wave observatories.
    Dashed black lines in the left-hand plot indicate the expected magnitude of
    several important backgrounds,
    in particular super-massive black hole binaries (SMBHB)~\cite{Simon:2016ibt,
    Arzoumanian:2018saf}, and galactic~\cite{Cornish:2017vip, Cornish:2018dyw} as
    well as extra-galactic~\cite{Farmer:2003pa, Yagi:2011yu} compact binaries (CB).
    In determining the power-law integrated sensitivity curves (as well as in
    the toy model analyses presented in \cref{sec:toymodels}), we assume that
    the SMBHB background will eventually be resolvable, while the CB background
    will remain unresolved. In the right-hand plot, we also show example
    spectra generated by a phase transition at $T^\text{nuc}=\SI{10}{GeV}$ and
    with $\alpha=0.1$, $\beta/H=10$ for both runaway and non-runaway bubbles.
    The parameter choices made for forthcoming experiments are given in
    \cref{sec:appendix-sensitivity}, and the data underlying our noise curves
    and PLI sensitivity curves can be found in the ancillary material.
  }
  \label{fig:noise_and_power-law-int-sense}
\end{figure}

To investigate the detectability of the predicted gravitational wave
signals from hidden sector phase transitions, we follow a frequentist
approach by calculating the corresponding signal-to-noise ratio (SNR) \SNR.
A stochastic gravitational wave background is detectable if the signal-to-noise is greater
than a certain threshold value $\SNR_\text{thr}$, which is either given
by the experimental collaborations or extracted from existing data as described
in \cref{sec:appendix-sensitivity}.

The optimal-filter cross-correlated signal-to-noise is~\cite{Thrane:2013oya,
Caprini:2018mtu}\footnote{For the case of a single-detector auto-correlated
  analysis, the factor $2$ in \cref{eq:Sensitivity-crosscorrelatedSNR} has to
be dropped.}
\begin{align}
  \SNR^2 = 2 \,\Tobs \int\limits_{f_\text{min}}^{f_\text{max}} \! \upd\!f \,
                       \left[ \frac{\hhOmega{\text{GW}}(f)}
                                   {\hhOmega{\text{eff}}(f)} \right]^2 \,,
  \label{eq:Sensitivity-crosscorrelatedSNR}
\end{align}
where \Tobs is the duration of the observation,
$\ba{f_\text{min},f_\text{max}}$ is the detector frequency band, and
$\hhOmega{\text{eff}}(f)$ is the effective noise energy density, i.e.\ the noise
spectrum expressed in the same units as the spectral gravitational wave energy
density~\cite{Thrane:2013oya}.  See \cref{sec:appendix-SNR} for more details. 

To make the comparison between the predicted signal and the noise even simpler,
it has become standard practice to quote so-called power-law integrated (PLI)
sensitivity curves~\cite{Thrane:2013oya}.  They are obtained by assuming the
gravitational wave spectrum follows a power law with spectral index $b$, i.e.\
\begin{align}
  \hhOmega{\text{GW}}(f) = \hhOmega{b} \, \bigg(\frac{f}{\bar{f}} \bigg)^b \,,
  \label{eq:Sensitivity-PLSpectra}
\end{align}
where $\hhOmega{b}$ is the gravitational wave energy density at the arbitrarily chosen
reference frequency $\bar{f}$.  According to \cref{eq:Sensitivity-crosscorrelatedSNR},
such a power-law signal is detectable if
\begin{align}
  \hhOmega{b} > \hhOmega{b}^\text{thr} \equiv
                   \frac{\SNR_\text{thr}}{\sqrt{2\Tobs}}
                   \bb{\, \int\limits_{f_\text{min}}^{f_\text{max}} \! \upd f \,
                          \ba{ \frac{\ba{f/\bar f}^b}{\hhOmega{\text{eff}}(f)} }^2
                   }^{-\frac{1}{2}} \,.
  \label{eq:PLI-h2Omega}
\end{align}
The PLI sensitivity curve is then obtained by determining
$\hhOmega{b}^\text{thr}$ as a function of the spectral index $b$,
and quoting the envelope of the corresponding power-law spectra as the sensitivity
limit of the experiment. In other words, the PLI sensitivity curve is given by
\begin{align}
  \hhOmega{\text{PLI}}(f) = \max\limits_b \bigg[ \hhOmega{b}^\text{thr}
                                                     \bigg( \frac{f}{\bar f}\bigg)^b
                                          \bigg] \,.
\end{align}
The region enclosed by the PLI sensitivity curve can be interpreted
as the region accessible by the experiment. Pictorially, a gravitational wave
spectrum
that reaches into this region has a sufficiently large signal-to-noise ratio to be detected.
Strictly speaking, this interpretation is only true for spectra that follow a
power-law in the frequency band of
the detector. However, realistic stochastic gravitational wave backgrounds are at least
approximate broken power laws.  This means in particular that at the
points where the predicted spectrum crosses the sensitivity limit,
it is usually well-described locally by a power-law, making the comparison to
the PLI sensitivity curve meaningful.
Nevertheless we will use \cref{eq:Sensitivity-crosscorrelatedSNR} directly to evaluate the
detectability of specific sets of model parameters.

The noise levels and PLI sensitivity curves for a number of gravitational
wave observatories are shown in \cref{fig:noise_and_power-law-int-sense}.
In particular, we consider the proposed ground-based Einstein
Telescope~(ET)~\cite{Sathyaprakash:2012jk}, the planned space-based
LISA~\cite{Audley:2017drz} interferometer as well as the proposed successor
experiments BBO~\cite{Crowder:2005nr} and (B-)DECIGO~\cite{Seto:2001qf,Sato:2017dkf}).
Moreover, we include PTAs, in particular the
currently operating EPTA~\cite{Lentati:2015qwp} and NANOGrav (NG) \cite{Arzoumanian:2018saf}, as well as the future SKA~\cite{Janssen:2014dka} telescope.
The data underlying these curves are included in machine-readable form as
ancillary material.

\subsection{Results}
\label{sec:sens-results}

\begin{figure}
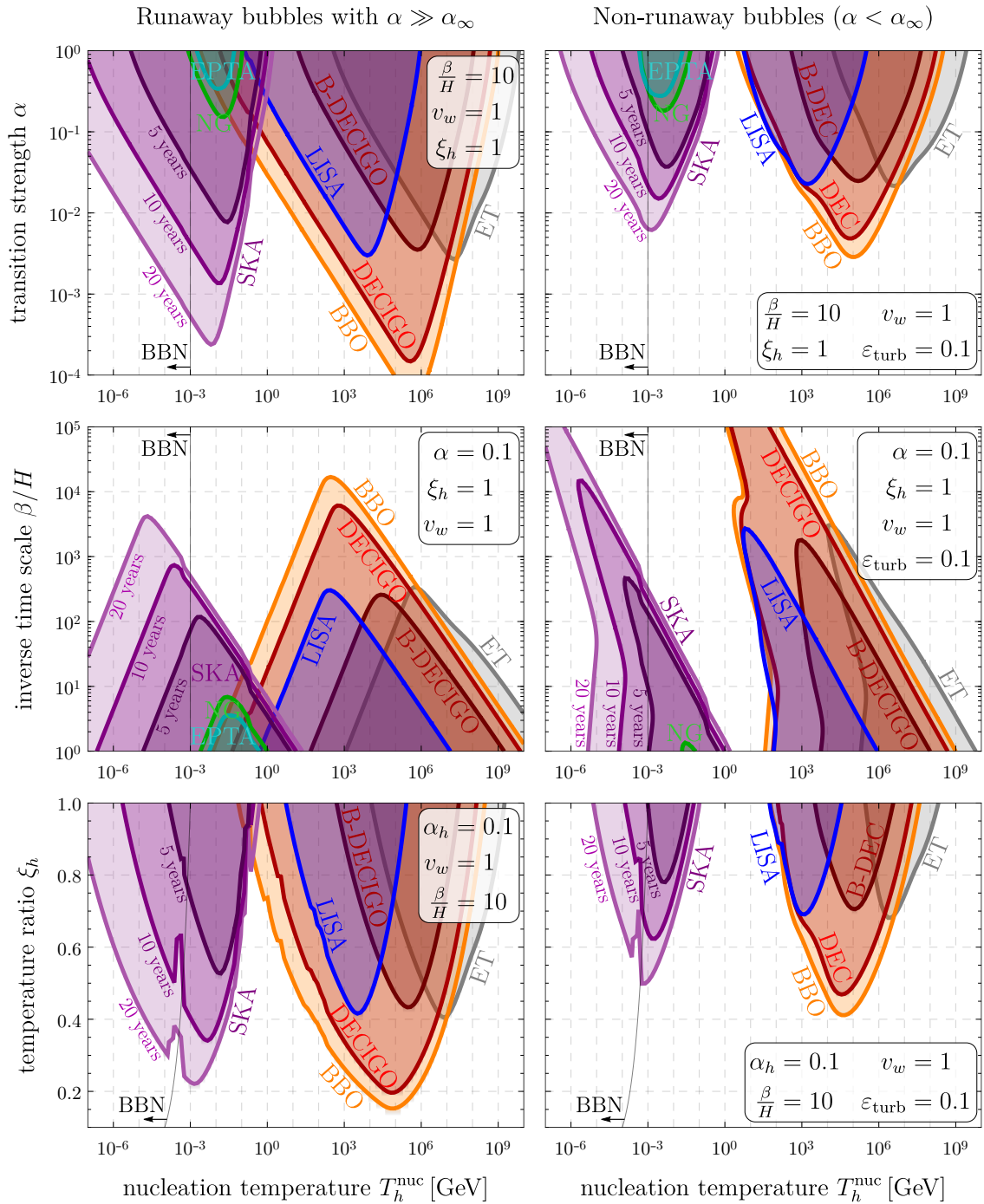

    \centering
    \includegraphics[width=\textwidth, trim={1.65cm 1.2cm 1.65cm 0.4cm}, clip]
      {{{sensitive_regions/T-alpha_beta=10}}}\\
    \includegraphics[width=\textwidth, trim={1.65cm 1.2cm 1.65cm 1cm}, clip]
      {{{sensitive_regions/T-beta_alpha=0.1}}}\\
    \includegraphics[width=\textwidth, trim={1.65cm 0.5cm 1.65cm 1cm}, clip]
      {{{sensitive_regions/T-rT_alpha=0.1_beta=10}}}
    \caption{Anticipated sensitivity to hidden sector phase transitions for
      various future gravitational wave observatories. In the left-hand panels
      we assume runaway bubbles, while the right-hand panels show the case of
      non-runaway bubbles. We show the sensitivities as a function of the
      hidden sector temperature at which bubble nucleation occurs, $\Thnuc$,
      versus the transition strength $\alpha$ (top), the inverse time scale
      $\beta/H$ (middle), and the temperature ratio between the hidden and visible
      sector $\temprat{h}$ (bottom). In all panels, we have assumed $\DOFh \ll
      \gG$ in calculating the redshifting of gravitational wave spectra.
      Note that in the bottom panel, we fix
      $\alpha_h$ (the value of $\alpha$ at $\temprat{h} = 1$) instead of
      $\alpha$ to explicitly show the $\temprat{h}$-dependence of $\alpha$ for
      fixed values of \Thnuc. Note that the translation of $\alpha_h$ to the
      physical $\alpha$ also relies upon the assumption $\DOFh \ll
      \gG$. \cref{fig:max_hidden_dofs} shows the resulting sensitivity when this
      assumption is relaxed. The discontinuities in
      the lower panel (best
      visible for SKA) originate from a step-function approximation to $\gG$.
    }
    \label{fig:sensitive-regions}
\end{figure}
\begin{figure}
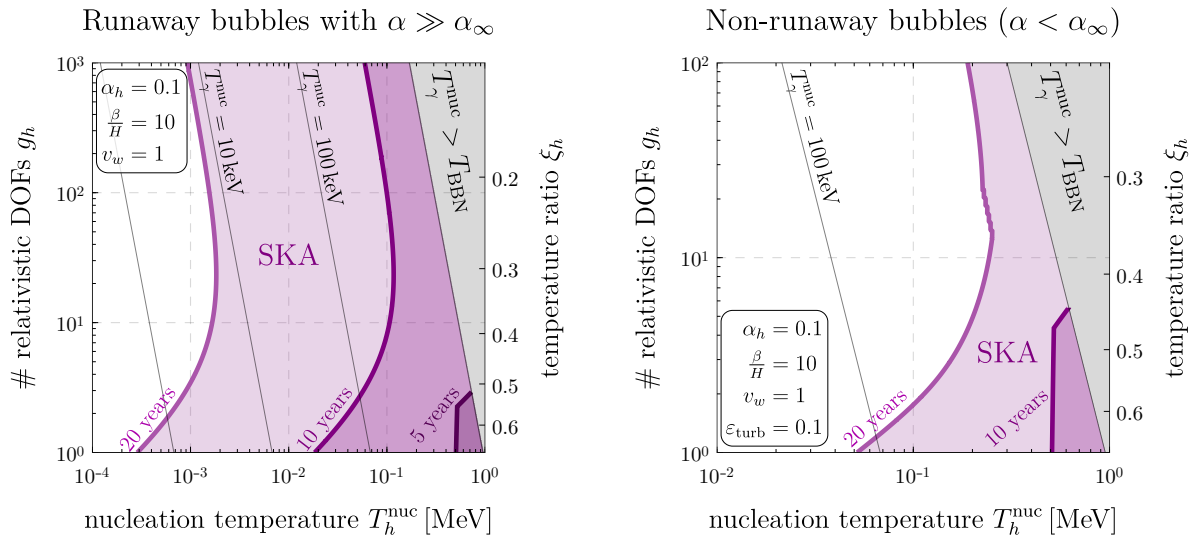

  \includegraphics[width=0.475\textwidth]{{{max_hidden_dof/maxhiddendof_runaway}}}
  \hspace*{0.024\textwidth}
  \includegraphics[width=0.475\textwidth]{{{max_hidden_dof/maxhiddendof_nonrunaway}}}
  \caption{Requirements for observable gravitational wave signals from a phase transition
    in a fully decoupled hidden sector in terms of the nucleation temperature and the
    number of relativistic degrees of freedom at BBN.  Models within the purple
    shaded region saturate the BBN constraints on \Neff (see \cref{eq:Neff-dec})
    and yield a gravitational wave signal observable in SKA.  The left-hand (right-hand)
    panel corresponds to runaway (non-runaway) bubble walls.
    The gray shaded region
    indicates that the phase transition occurs before the onset of BBN,
    alleviating any constraints on the maximal number of degrees of freedom.
  }
  \label{fig:max_hidden_dofs}
\end{figure}

We now translate the noise curves from \cref{fig:noise_and_power-law-int-sense}
(left) into anticipated constraints on the phase transition parameters
$\alpha$, $\beta$, and $\Thnuc$ (introduced in \cref{sec:gw-params}) and on the
temperature ratio $\temprat{h}$.  In \cref{fig:sensitive-regions}, we show the
regions where $\rho > \rho_\text{thr}$ as a function of $\Thnuc=\Tnuc$ and
$\alpha$ (top), $\Thnuc=\Tnuc$ and $\beta/H$ (middle), and $\Thnuc$ and
$\temprat{h}$ (bottom). The left-hand (right-hand) panel corresponds to runaway
(non-runaway) bubble walls.  To show the $\temprat{h}$ dependence for fixed
values of \Thnuc in the bottom panel, we do not fix $\alpha$ (which itself
depends on $\temprat{h}$).  Instead, we keep the latent heat fixed by fixing
the parameter $\alpha_h$ defined as the value of $\alpha$ assuming $\temprat{h}
= 1$. 

The upper panels of \cref{fig:sensitive-regions} show that PTAs
have optimal sensitivity to phase transitions occurring at hidden sector
temperatures between $\SI{1}{\keV}$ and $\SI{1} {\GeV}$ for non-runaway
bubbles, with the lower reach increasing far below the $\SI{}{\keV}$ scale for
runaway bubbles.  Ground-based and space-based interferometers will cover the
range from $\SI{10}{GeV}$ to $\order{\si{PeV}}$, where again the lower reach
increases to the \SI{} {\MeV} scale for non-runaway bubble walls.  As expected,
the discovery reach is best for large $\alpha$ and small $\beta/H$,
corresponding to phase transitions that involve highly energetic collisions of
large bubbles.  The middle right-hand panel of \cref{fig:sensitive-regions}
reveals that the high temperature boundary of the sensitivity curve shifts
towards lower temperature scales for increasing $\beta/H$. As a consequence,
fast phase transitions ($\beta/H \gg 1$) will be detectable by pulsar timing
only for $\Thnuc\lesssim \si{MeV}$. A mass spectrum at that scale and below can
easily come into conflict with cosmological constraints.

The characteristic kink in the low temperature boundary of the sensitivity
curve for $\beta/H$ versus \Thnuc and non-runaway bubbles (middle right-hand
panel of \cref{fig:sensitive-regions}) arises due to the shape of the
gravitational wave spectrum. For non-runaway bubbles the spectrum comprises of
two contributions (sound wave and turbulence), compared to the single dominant
piece for the runaway scenario with $\alpha \gg \alpha_\infty$.  Starting with
small $\beta/H$ at the lower edge of the sensitivity curve, the peak frequency
for both contributions lies at frequencies below the lower frequency threshold
of the detector. However, due to the large amplitude ($\alpha=0.1$), there is
nevertheless sensitivity to the high-frequency tail of the turbulence
contribution. Increasing $\beta/H$ decreases the amplitude and increases the
peak frequency of both contributions.  Thus, the spectral tail of the
turbulence contribution drops below the experimental sensitivity, but
simultaneously, the peak region of the spectrum (which is dominated by sound
waves) moves into the sensitivity region.  The kink in the middle right-hand
panel of \cref{fig:sensitive-regions} indicates where the sensitivity changes
from being dominated by turbulence to being dominated by sound waves.  In
contrast, the case of runaway bubbles results in a spectrum dominated by the
scalar field contribution, leading to sensitivity curves that are roughly
symmetric.

For values of $\beta/H \gtrsim 100$ the amplitude of the gravitational wave spectrum arising from sound waves caused by non-runaway bubbles may be overestimated. In this regime the sound wave contributions are expected to last less than a Hubble time, quickly transitioning to turbulent flows in the plasma. This will result in a decrease in amplitude of the gravitational waves sourced from sound waves, but is argued to increase the contribution arising from turbulence \cite{Hindmarsh:2017gnf,Ellis:2018mja}. Such an increase is expected to mitigate the overall reduction in the amplitude of the gravitational wave spectrum. Given that this transition between the two regimes cannot yet be accurately simulated, the exact decrease in magnitude remains unknown. Hence, we neglect these dynamics in both \cref{fig:sensitive-regions} and the model dependent results in the following section.

Comparing the sensitivity to phase transitions with runaway bubbles
(left column in \cref{fig:sensitive-regions}) with the sensitivity
to non-runaway transitions (right column), we see that in the runaway case,
significantly larger regions of parameter space can be probed. The
only
exception arises for large values of $\beta/H$. From \cref{tab:gw-spectra} we
observe that
the scalar field contribution is suppressed by an extra power of $\beta/H$
compared to both the sound wave and turbulence contributions.

In the bottom row of \cref{fig:sensitive-regions}, we observe that the
sensitivity drops for $\temprat{h} < 1$, which is due to the scaling of the
transition strength parameter $\alpha \propto \temprat{h}^{4}$ for fixed \Thnuc (see
\cref{eq:alpha,eq:rho-R}).  For small $\temprat{h}$, the photon temperature
$\Tnuc$ at the time of the phase transition is much larger, hence the $
\temprat{h}^4$ suppression. We emphasize here that for a fixed value of
the latent
heat (or equivalently for a fixed $\alpha_h$, the strength of the transition assuming $\temprat{h}=1$) a strong phase transition is more easily obtained at later
times when the radiation energy density of the universe is lower. This increase
is then offset by the effects of the temperature ratio, with the net result
being an observable gravitational wave signal for large regions of $\alpha$
and $\beta/H$.

In \cref{fig:max_hidden_dofs} we show the maximum number of degrees of freedom
that can be present in a hidden sector whilst being consistent with \Neff
measurements at BBN and simultaneously producing an observable gravitational
wave signal in SKA. This figure is produced assuming the number of degrees of
freedom saturates the BBN constraint shown in the left-hand panel of
\cref{fig:Neff-constraints}, i.e.\ we assume that the hidden sector is
decoupled from the visible sector and remains so throughout the thermal
history. (In the $\nu$\mbox{-}quilibration scenario, there would be an additional
parameter $\tempratinit$, making visualization difficult.) We see that in the
runaway case, SKA may be sensitive to hidden sectors with an $\order{1}$ number
of relativistic degrees of freedom already after 5~years of observation time.
Discovery prospects are best for a phase transition at the beginning of BBN. In
the non-runaway scenario, at least 10~years of observation time are required to
access models with an $\order{1}$ number of relativistic degrees of freedom.
For longer observation periods, the reach will include models with significantly
more degrees of freedom within certain temperature ranges.  The shape of the
sensitivity curves in \cref{fig:max_hidden_dofs} can be understood by noting
that, even for arbitrarily large $\DOFh$, the $\Neff$ constraint forces us to
keep the total radiation energy density fixed.  This implies that a change in
$\DOFh$ is accompanied by a compensating change in $\temprat{h}$.  Since we
have chosen a fixed $\alpha_h$ and $\beta/H$, this means that a larger number
of degrees of freedom simply implies that the phase transition happens earlier
($\Tnuc \propto \DOFh^{1/4}$), that $\alpha$ decreases $\propto \DOFh^{1/4}$
(see \cref{eq:alpha}), and that the peak frequency increases $\propto
\DOFh^{1/4}$ (see \cref{tab:gw-spectra}).  Combining this behavior with the
shape of the SKA sensitivity curve from
\cref{fig:noise_and_power-law-int-sense} leads to the sensitivity regions shown
in \cref{fig:max_hidden_dofs}.

We emphasize again that we have here assumed that the stochastic
SMBHB background will eventually be resolvable. If this is not the case,
the sensitivity regions of SKA shrink almost to the size of the EPTA and
NANOGrav regions.

\section{Models}
\label{sec:toymodels}

We now apply the model-independent results from \cref{sec:sensitivity} to
two specific toy models, which can serve as benchmarks for future studies. We
focus on models that feature first-order hidden sector phase transitions at
scales $\lesssim \text{MeV}$ and might thus be detectable by PTAs.  In addition we focus purely on the hidden sector field content and symmetries, setting potentially allowed portal couplings to the SM exactly to zero.  This is motivated by the requirement that any portal coupling present must be sufficiently small\footnote{In the case of a scalar portal, forbidding thermal equilibrium through scattering all the way down to the \SI{}{\MeV} scale requires $\lambda \lesssim \SI{E-11}{}$.} to forbid thermal equilibrium between the two sectors. A detailed discussion of concrete realizations yielding a temperature ratio between the two sectors as well as the allowed coupling values that preserve it until the phase transition is given in Ref.~\cite{Fairbairn:2019xog}. Finally, because of the strong \Neff constraint, any ultraviolet-complete model with non-trivial
dynamics at the MeV scale should reduce to one of a small number of effective
low energy models.

\subsection{Singlet Scalars}
\label{sec:singlet-scalar}

\begin{figure}
    \includegraphics[width=0.475\textwidth]{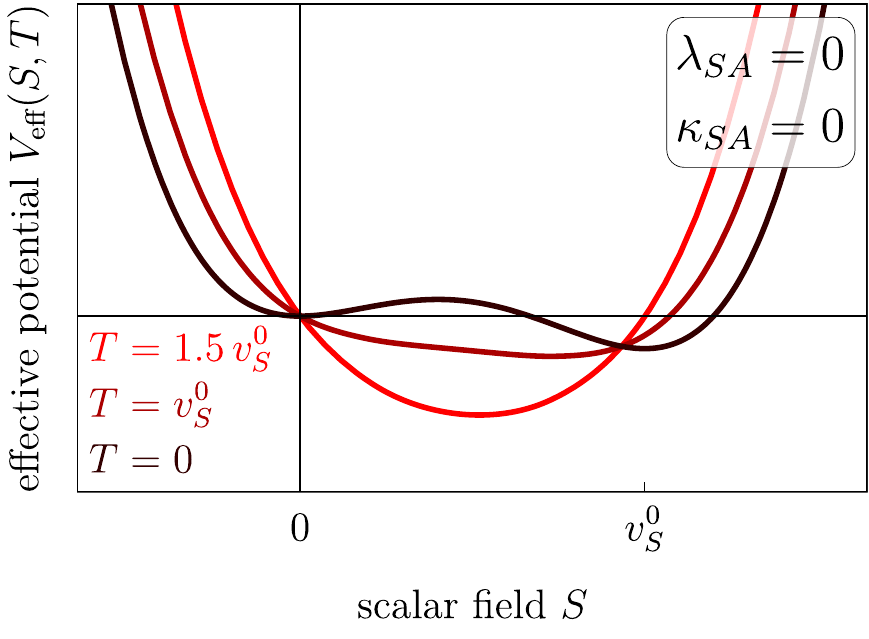}
    \hspace*{0.024\textwidth}
    \includegraphics[width=0.475\textwidth]{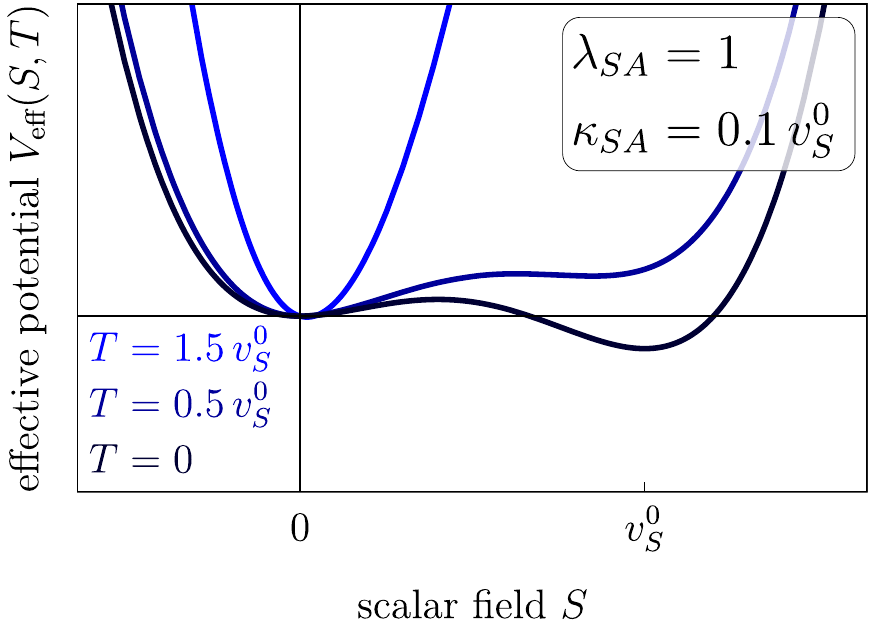}
    \caption{Qualitative behavior of the effective potential $V_\text{eff}
      (S,T)\simeq V_\text{tree}(S)+V_T(S,T)$ in the singlet scalar model
      without an auxiliary scalar field $A$ ($\lambda_{SA}=\kappa_{SA}=0$, left panel)
      and with the inclusion of such a field ($\lambda_{SA},\kappa_{SA}>0$, right plot).
      Only in the latter case is a first-order phase transition
      realized. For illustration purposes we have shifted the potentials such
      that $V_\text{eff}(0,T)=0$.}
    \label{fig:singletscalar-potential}
\end{figure}

For our first toy model, we consider a hidden sector in which a real scalar
singlet $S$ acquires a tree-level potential barrier between the two possible
minima of its scalar potential.  In the most minimalistic setup,
such a barrier is generated by the cubic term $\propto \kappa S^3$ in the
potential.  If this coupling is non-zero, and if the mass term $\frac{1}{2}
\mu_S^2 S^2$ is positive, the tree-level potential at zero temperature has one
minimum at $v_S \equiv \ev{S} = 0$, and a second one away from the origin. The
effective potential at high temperatures, on the other hand, has only a single
minimum.  This can be seen from both the left-hand panel of
\cref{fig:singletscalar-potential} and from the analytical form of the
high-temperature expansion (\cref{eq:VT-S} in
\cref{sec:appendix-effpot}).\footnote{Here and in the following, we denote by
$S$ an arbitrary value of the field, while we reserve the notation $v_S$ (or,
equivalently, $\ev{S}$) for the value of the field at its global minimum.}
The leading term in this effective potential is $V_\text{eff}(S,T) \simeq
V_T(S) \propto m_S^2(S) T^2$, where $m_S^2(S)$ is the field-dependent mass
parameter, i.e.\ the second derivative of the tree-level potential with
respect to $S$.  From its explicit form given in \cref{eq:mS2}, we see that
$V_T(S)$ is approximately parabolic at high $T$, with a minimum shifted away
from the origin. Following the evolution of $V_\text{eff}(S,T)$ to lower
temperatures, we find that this minimum evolves smoothly into the global
minimum of the zero-temperature potential (see left-hand panel of
\cref{fig:singletscalar-potential}). This means that the field never finds
itself in a false vacuum, and no first-order phase transition occurs.

The situation is different if we introduce a second, auxiliary, real singlet
scalar $A$.  The resulting scalar potential is given by
\begin{align}
  V_\text{tree}(S,A)
      &= \frac{\mu_S^2}{2} S^2 
       + \frac{\kappa}{3} S^3 
       + \frac{\lambda_S}{4}S^4 
       + \frac{\mu_A^2}{2} A^2 
       + \frac{\lambda_A}{4} A^4
       + \kappa_{SA} S A^2 
       + \frac{\lambda_{SA}}{2} S^2A^2 \,,
  \label{eq:Vtree-scalar}
\end{align}
where we have assumed that $A$ is odd under a $\mathbb Z_2$ symmetry.  We
furthermore assume $\mu_A^2, \kappa_{SA}\ge 0$ to
avoid spontaneous breaking of the $\mathbb Z_2$ symmetry.  The scalar quartic
couplings $\lambda_S$, $\lambda_A$ are required to be positive to
ensure stability of the potential. The quartic and trilinear portal terms, $\frac{1}{2} \lambda_{SA} S^2 A^2$ and $\kappa_{SA}SA^2$, contribute quadratically and linearly through the thermal potential $V_T$. These contributions shift the minimum in the $S$-field direction closer to the origin at high temperatures. This enables $S$ to
first evolve into the false minimum at the origin as the temperature drops, and
eventually tunnel into the true minimum in a first-order phase transition 
(as illustrated in the right panel of \cref{fig:singletscalar-potential}).
In particular, an analysis of the tree-level potential implies that such a transition occurs if
\begin{align}\label{eq:kappabardef}
  \bar\kappa \equiv -\frac{\kappa}{\lambda_S v_S^0}
             \in \big(1 \dots \tfrac{3}{2} \big) \,,
\end{align}
where $v_S^0 \equiv v_S(T=0)$ is the vacuum expectation value (vev)
of $S$ at zero temperature.
In the following, we will treat $v_S^0$ as an input parameter of the model
and solve the extremal condition $\partial_S V_\text{tree}(S, 0)|_{S=v_S^0} = 0$ to
determine $\mu_S$.

\begin{figure}
  \centering
  \includegraphics[width=0.495\textwidth,trim={0cm 0cm 0cm 0cm}]{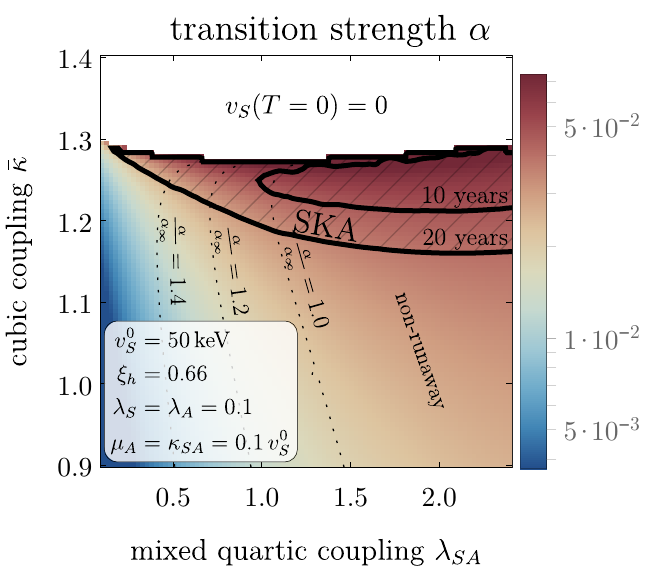}
  \includegraphics[width=0.495\textwidth,trim={0cm 0cm 0cm 0cm}]{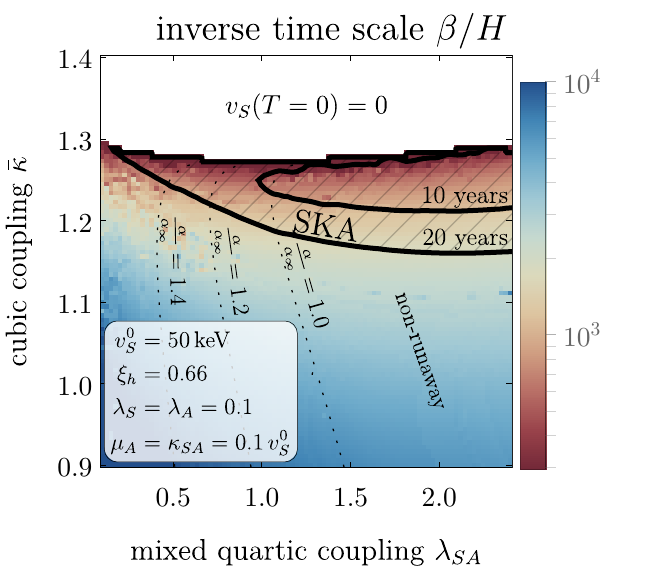}
  \caption{The strength of the hidden sector phase transitions $\alpha$ (left panel)
    and its inverse time scale $\beta/H$ (right panel) for the singlet scalar
    model given by \cref{eq:Vtree-scalar} at the scale $\Thnuc\sim v_S^0 =
    \SI{50}{keV}$. Inside the regions bounded by the black
    lines, the gravitational wave signals from the phase transition will be
    detectable by SKA after the indicated periods of observation time. To
    satisfy
    the cosmology constraints for $v_S^0 =
    \SI{50}{keV}$, we require the hidden sector
    to be colder than the visible sector by a factor of $\temprat{h} = 0.66$
    at the time of the transition. This is the
    temperature ratio that arises naturally in our $\nu$\mbox{-}quilibration scenario (see \cref{eq:xi-reeq-1,eq:Neff-reeq-2}), satisfying the CMB+$H_0$ cosmology constraint
    given in \cref{eq:Neff-constraint-CMB-H0}.
  }
  \label{fig:scalarsinglets-scan}
\end{figure}

\Cref{fig:scalarsinglets-scan} shows the resulting gravitational wave
parameters $\alpha$ and $\beta/H$ in the $\bar\kappa$--$\lambda_{SA}$ plane at
$v_S^0 = \SI{50}{keV}$. To obtain these plots, we have simulated the phase
transition using the \texttt{CosmoTransitions} package~\cite{Wainwright:2011kj,
Kozaczuk:2014kva, Blinov:2015sna, Kozaczuk:2015owa}, which we have extended to
compute gravitational wave parameters and to improve the stability and
performance during parameter scans.  The actual gravitational wave spectra were
then taken from the hydrodynamic simulations discussed and referenced in
\cref{sec:gw-spectrum}.\footnote{As we are including both the sound wave and
turbulence contributions to the gravitational wave spectrum, the results here
rely on bubble nucleation occurring in a plasma. However, in contrast to the
SM plasma, before and after the phase transition
there are no completely massless degrees of
freedom in the model.  Nevertheless, the hierarchy of the mass spectrum in
relation to the nucleation temperature ensures that there remains at least one
relativistic degree of freedom and therefore a plasma in the broken phase of
the hidden sector. The time frame before these relativistic degrees of freedom
become non-relativistic is sufficient for the sound wave contributions to
source gravitational waves as the scenarios considered here have large
$\beta/H$ values and negligible supercooling. A similar picture holds for the dark
photon model considered in the following section.} Details on the computation
of the effective potential that is used as an input for
\texttt{CosmoTransitions} are given in \cref{sec:appendix-effpot}.  A
first-order transition occurs only in the shaded regions of parameter space.
The parameters $\bar\kappa$ and $\lambda_{SA}$ are the most important handles
controlling the dynamics of the phase transition. As already mentioned,
$\lambda_{SA}$ (or a combination of $\lambda_{SA}$ and $\kappa_{SA}$ in the case of a non-zero cubic) is required to trap the high-temperature vacuum in a local
minimum. This explains why increasing $\lambda_{SA}$ opens up more parameter
space in which a first-order transition occurs. $\bar\kappa$ on the other hand
controls the size of the potential barrier between the true and false vacua.
The transition becomes slower and more energetic as $\bar\kappa$ is increased.
Above a certain threshold, however, the nucleation criterion in
\cref{eq:nuc-cond} is never met (i.e.\ the tunneling action never drops below
the threshold value required for the transition to begin). In this case, the
field remains trapped in the false minimum with $v_S = 0$.  The upper bound on
$\bar\kappa$ quoted above is $\bar\kappa < \frac{3}{2}$ at tree
level, however this upper bound is further reduced when
radiative and temperature-dependent corrections
are included.  Note also that the bubbles of the phase transitions in
the parameter region of interest are close to the boundary between runaway and
non-runaway regime, i.e.\ $\alpha \simeq \alpha_\infty$, as indicated by the
dotted lines in \cref{fig:scalarsinglets-scan}.

\begin{figure}
	\makebox[\textwidth][c]{
		\includegraphics[width=\textwidth,trim={0.6cm 0.35cm 0.6cm 0.35cm},clip]
		{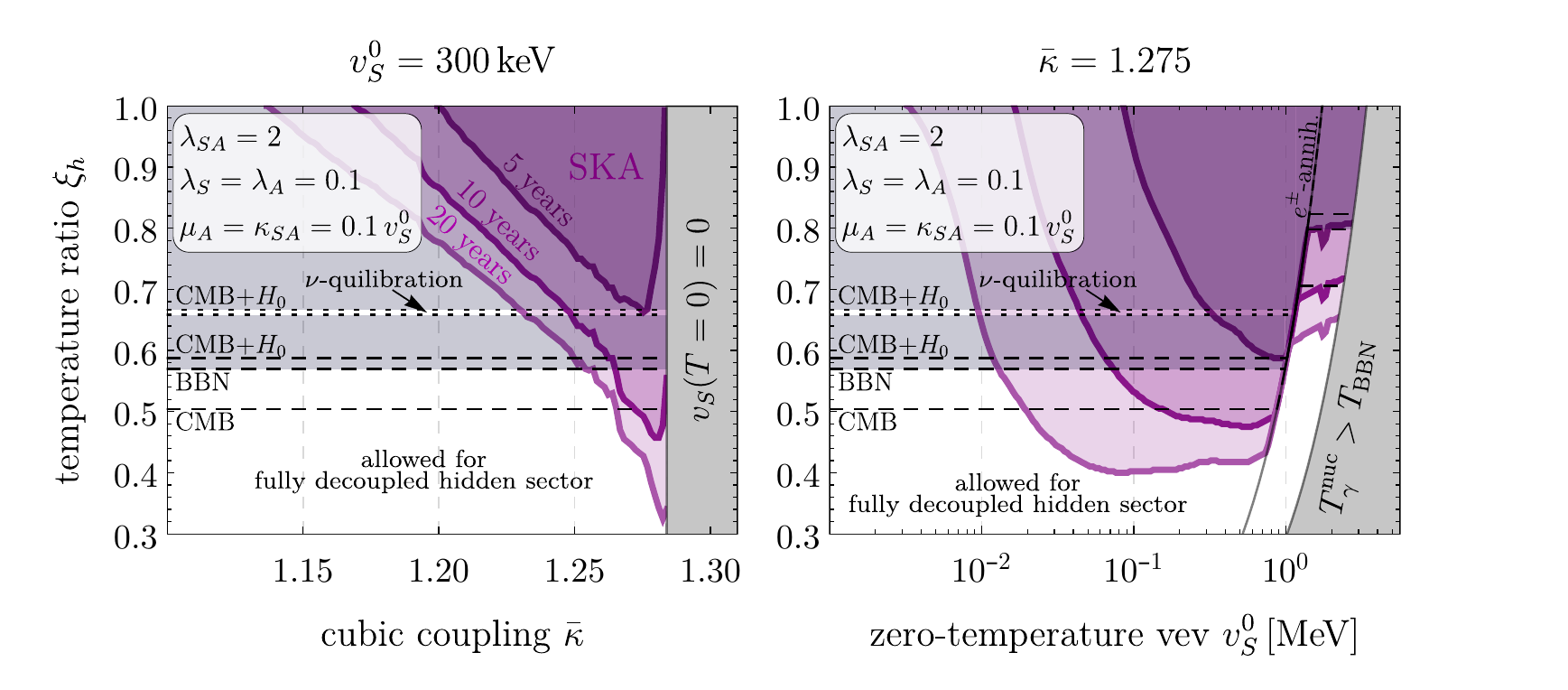}}
	\caption{
		Impact of the temperature ratio \temprat{h} at the time of the hidden
		sector phase transition on the sensitivity of SKA to the singlet scalar
		model from \cref{eq:Vtree-scalar}. The left panel shows the accessible
		values of $\temprat{h}$ as a function of $\bar\kappa$ (defined in
		\cref{eq:kappabardef}), while the right panel shows them as a function of
		the zero-temperature vev $v_S^0$. The gray region above the horizontal
		dashed lines indicates the values of $\temprat{h}$ excluded by the CMB, CMB+$H_0$ and BBN constraints on \Neff (see
		\cref{eq:Neff-constraint-CMB,eq:Neff-constraint-BBN,eq:Neff-constraint-CMB-H0}) for a fully decoupled
		hidden sector. Note that the CMB constraints have been derived under the
		assumption that the hidden sector energy density redshifts as radiation. In
		realistic scenarios, hidden sector particles will eventually become
		non-relativistic before decaying or annihilating to radiation.  This will
		tighten the CMB constraints (i.e.\ shrink the white region) in a
		model-dependent manner. The narrow white band delimited by dotted black
		lines corresponds to the range of $\temprat{h}$ accessible in the
		$\nu$\mbox{-}quilibration scenario, in which the hidden sector equilibrates with
		the SM neutrinos after the latter have decoupled. Its width is given by
		varying $\tempratinit$ between zero and the maximum value allowed by the
		CMB$+H_0$ constraint, $\tempratinit \simeq 0.61$. The $\nu$\mbox{-}quilibration
		scenario is only allowed by the less stringent CMB$+H_0$ bound on \Neff,
		but not by the CMB-only constraint. Note that $\Thnuc\simeq 0.3 v_S^0$ for
		the parameter regions shown above.
	}
	\label{fig:singletscalars-xi}
\end{figure}

In \cref{fig:singletscalars-xi} we show the expected sensitivity of SKA to the
gravitational wave signal from the phase transition in the scalar singlet
model.  We vary $\bar\kappa$ and $v_S^0$, while keeping all dimensionless
couplings fixed at $\order{1}$ values, as indicated in the plots. We choose the remaining dimensionful parameters $\mu_A$ and $\kappa_{SA}$ such that $A$ is heavier than $S$ while their zero-temperature masses are both of order $v_S^0$. This is a requirement of the $\nu$-equilibration scenario such that $A$ doesn't remain as a stable thermal relic, while the lighter field $S$ decays to SM neutrinos. A concrete incarnation of this mechanism is discussed at the end of this subsection.  We vary the temperature ratio $\temprat{h}$
between the visible and hidden sector at the time of the phase transition
(vertical axis), indicating which parameter regions are allowed by cosmological
bounds on \Neff (white areas).  If the hidden sector remains decoupled
throughout the post-BBN evolution of the Universe, $\temprat{h} \lesssim 0.50$, $\lesssim 0.59$ or $\lesssim 0.57$ is required by the CMB, CMB+$H_0$ and BBN constraint given in
\cref{eq:Neff-constraint-CMB}, \cref{eq:Neff-constraint-CMB-H0} and \cref{eq:Neff-constraint-BBN} respectively. If the hidden
sector equilibrates with neutrinos before the phase transition, then the
temperature ratio during the phase transition is $\temprat{h} \sim
0.66$.  The boundaries of the corresponding narrow white band are determined by
varying the initial temperature ratio before (re\mbox{-})coupling between $\tempratinit
= 0$ and the maximum possible value $\tempratinit \simeq 0.61$ from the
CMB$+H_0$ bound of \cref{fig:Neff-constraints}.

In this context, let us comment on possible mechanisms to realize
$\nu$\mbox{-}quilibration.  One possibility is to introduce a small Yukawa-like
coupling between $S$ and the light neutrinos \cite{Berlin:2017ftj,
Berlin:2018ztp}, for instance via a right-handed neutrino, $N$, with an
interaction term of the form $\lambda S \overline{N} N$. The assumption of a
type-I seesaw then yields a suppressed coupling of the desired form
$\frac{\lambda m_\nu}{m_N} S \overline{\nu} \nu$, where $\nu$ is a light
neutrino field after electroweak symmetry breaking, and $m_\nu$ and $m_N$ are
the masses of the light and heavy neutrinos, respectively.   While both $S$ and
the neutrinos are relativistic, the interaction rate between the hidden and
visible sectors scales proportional to $\Tg$, i.e.\ it drops more slowly than
the Hubble rate $H \sim \Tg^2 / M_\text{Pl}$ (where $M_\text{Pl}$ is the
Planck mass). Therefore, it will be initially smaller than $H$, but may
become larger at late times, post-BBN.  Once the hidden sector particles
become non-relativistic, their annihilation back to SM neutrinos can also
efficiently proceed as the lightest hidden sector particle, $S$, now has a
decay mode to SM neutrinos.

\subsection{Dark Photon}
\label{sec:dark-photon}

As a second toy model, we consider a scenario with an Abelian gauge symmetry
$U(1)'$ in the hidden sector. Gravitational wave signatures arising from
$U(1)'$ gauge symmetries broken at temperatures above
the MeV-scale have been considered before as part of a hidden sector 
\cite{Hashino:2018zsi,Addazi:2017gpt}, or as a thermally coupled extension of
the SM
gauge symmetries \cite{Madge:2018gfl}. Here, we focus instead on phase
transitions at temperatures below $\SI{1}{MeV}$.
We introduce a complex scalar $S$, which is a
singlet under the SM gauge groups, but charged under $U(1)'$. The relevant
terms in the Lagrangian are
\begin{align}
  \mathscr{L} &\supset
               \left|D_\mu S\right|^2 
               + \left|D_\mu H\right|^2 
               - \frac{1}{4}F_{\mu\nu}^\prime F^{\prime\,\mu\nu} 
               - \frac{\varepsilon}{2}F_{\mu\nu}^\prime F^{\mu\nu} 
               - V(S,H) \,,
  \label{eq:L-dark-photon}
\end{align}
where $F_{\mu\nu}^\prime$ and $F_{\mu\nu}$ are the field strength tensors of
$U(1)'$ and $U(1)_Y$, respectively, and $H$ is the SM Higgs field. The
covariant derivative acting on $S$ is
\begin{align}
  D_\mu S &= \left(\partial_\mu + i g_D A_\mu^\prime\right)S \,,
\end{align}
where $g_D$ is the coupling strength and $A_\mu^\prime$ is the gauge boson
of $U(1)'$. The most generic renormalizable scalar potential invariant
under the model's symmetries is given by
\begin{align}
  V_\text{tree}(S,H) &= - \mu_S^2 S^\dag S 
                        - \mu^2 H^\dag H 
                        + \frac{\lambda_S}{2} (S^\dag S)^2
                        + \frac{\lambda}{2} (H^\dag H)^2 
                        + \lambda_{SH} (S^\dag S)(H^\dag H) \,.
  \label{eq:Vtree-dark-photon}
\end{align}  
We see that the hidden and visible sectors can communicate with each other
through two portals: the kinetic mixing term between $U(1)$ gauge bosons,
and the mixed quartic (``Higgs portal'') coupling in the scalar potential.
For our purposes, we will assume that the dark sector is sequestered from
the visible sector, i.e.\ we set $\varepsilon =  \lambda_{SH} = 0$.  

\begin{figure}
	\centering
	\includegraphics[width=0.495\textwidth,trim={0cm 0cm 0cm 0cm}]{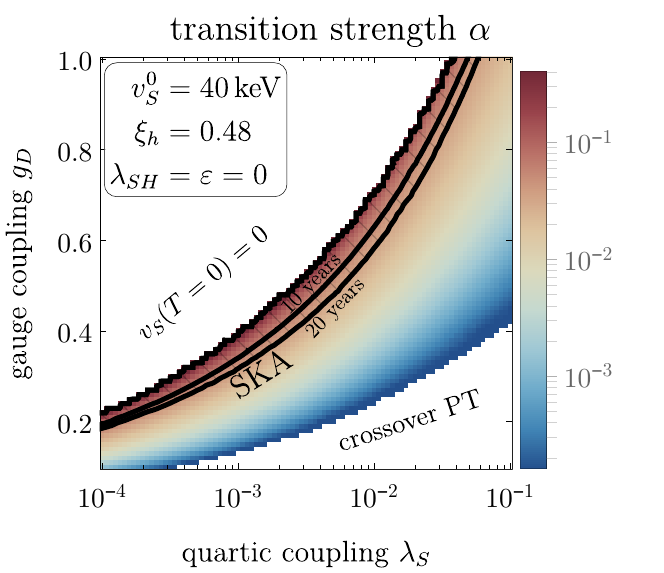}
	\includegraphics[width=0.495\textwidth,trim={0cm 0cm 0cm 0cm}]{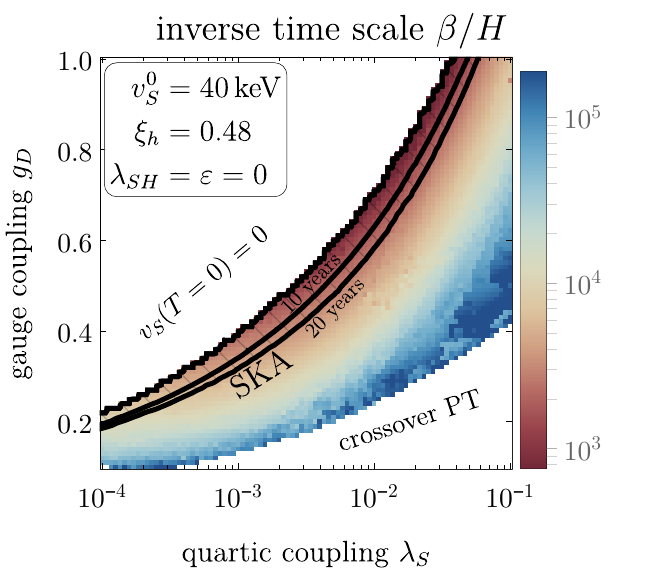}
	\caption{The strength of the hidden sector phase transition, $\alpha$, (left panel)
		and its inverse time scale $\beta/H$ (right panel) for the Higgsed dark
		photon model given by \cref{eq:L-dark-photon} at the scale $\Thnuc\sim
		v_S^0 = \SI{40}{keV}$. Inside the regions bounded by the black lines, the
		gravitational wave signals from the phase transition will be detectable by
		SKA after the indicated periods of observation time. To satisfy the
		cosmology constraints for $v_S^0 = \SI{40}{keV}$, we require the decoupled
		hidden sector to be colder than the visible sector by a factor of
		$\temprat{h} = 0.48$ at the time of the transition. This value satisfies
		the BBN constraint given in \cref{eq:Neff-constraint-BBN}.
	}
	\label{fig:darkphoton-scan}
\end{figure}

\Cref{fig:darkphoton-scan} shows our numerical results for the gravitational wave
parameters $\alpha$ and $\beta/H$ in the $g_D$--$\lambda_S$ plane at
$v_S^0 = \SI{40}{keV}$. Similar to $\bar{\kappa}$ in the singlet scalar
model, $g_D$ controls the size of the potential barrier in this case.
However, the barrier-inducing cubic term in the effective potential is now
a one-loop effect rather than tree level (see
\cref{sec:appendix-effpot} for details). For increasing values of $g_D$, the
transition
becomes slower and more energetic, until $g_D$ becomes too large and the scalar
vev remains trapped at $v_S = 0$. In the opposite limit of small $g_D$, the
transition becomes weaker and faster before entering the regime of smooth
crossover transitions without gravitational wave emission. With increasing
$\lambda_S$, the region in which a first-order transition occurs
deforms towards higher values of $g_D$. This is because, for fixed
$v_S^0$, the parameter $\mu_S^2$ which controls the depth of the potential
in the broken minimum is proportional to $\lambda_S$.
As a consequence, a deeper tree-level minimum (larger $\lambda_S$)
has to be paired with a larger barrier (larger $g_D$) in order to leave the
dynamics of the transition qualitatively unchanged. The tree-level masses of
the scalar $S$ and the dark photon $A'$ in the investigated parameter region
are $\lesssim v_S$ with $S$ being lighter than $A'$ because of the parameter
choice $g_D \lesssim 1$, $\lambda_S \ll 1$. Note that due to the additional
friction induced by the dark photon we assume non-runaway bubble walls in the
phase transition of this model~\cite{Bodeker:2017cim}.

\begin{figure}
	\makebox[\textwidth][c]{
		\includegraphics[width=\textwidth, trim={0.6cm 0.35cm 0.6cm 0.35cm}, clip]{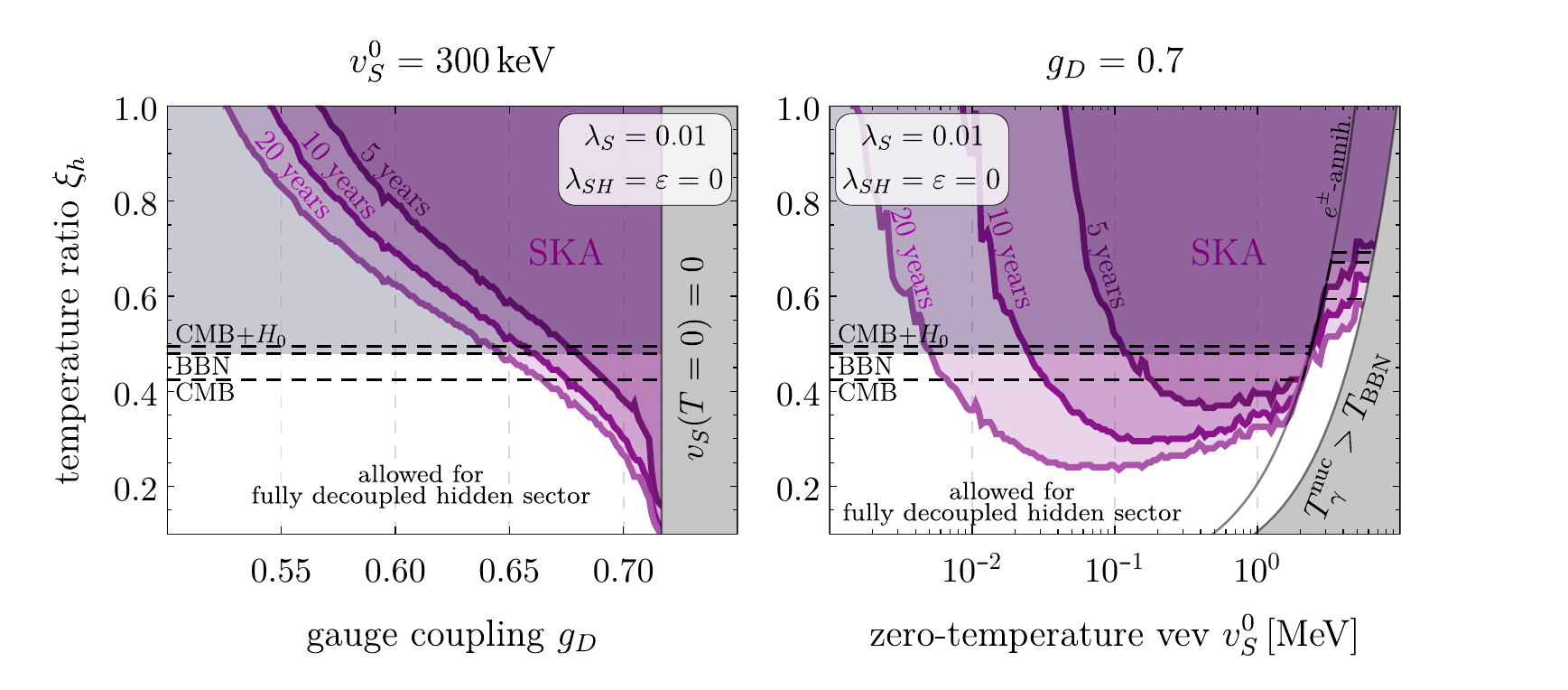}}
	\caption{Impact of the temperature ratio \temprat{h} at the time of the
		hidden sector phase transition on the sensitivity of SKA to the Higgsed dark
		photon model from \cref{eq:L-dark-photon}. The left panel shows the
		accessible values of $\temprat{h}$ as a function of $g_D$, while the right
		panel shows them as a function of the zero-temperature vev $v_S^0$. The gray region above  the horizontal
		dashed lines indicates the values of $\temprat{h}$ excluded by the CMB, CMB+$H_0$ and BBN constraints on \Neff (see
		\cref{eq:Neff-constraint-CMB,eq:Neff-constraint-BBN,eq:Neff-constraint-CMB-H0}) for a fully decoupled hidden sector. The
		$\nu$\mbox{-}quilibration scenario, in which the hidden sector (re\mbox{-})couples with the
		neutrinos after the latter have decoupled from the photons, has no allowed
		parameter space in this model.  Note that $\Thnuc\simeq 0.1 v_S^0$ for the
		parameter regions shown above.
	}
	\label{fig:darkphoton-xi}
\end{figure}

As for the singlet scalar model, we have also computed the sensitivity of SKA
to the parameter space of the dark photon model as a function of the temperature
ratio $\temprat{h}$. The results are shown in
\cref{fig:darkphoton-xi}.
Due to the additional degrees of freedom in the dark photon model compared to the singlet
scalars model, the scenario in which the hidden sector equilibrates with the neutrinos ($\nu$\mbox{-}quilibration) is completely excluded by the
CMB constraints on \Neff (see \cref{fig:Neff-constraints}). The white region indicates the allowed temperature ratios for a hidden sector that remains decoupled, corresponding to $\temprat{h}\lesssim 0.42$, $\lesssim 0.49$ and $\lesssim 0.48$ corresponding to the CMB, CMB+$H_0$ and BBN constraints given in \cref{eq:Neff-constraint-CMB}, \cref{eq:Neff-constraint-CMB-H0} and \cref{eq:Neff-constraint-BBN} respectively.

\subsection{Random Parameter Scans}
\label{sec:paramscans}

\begin{figure}
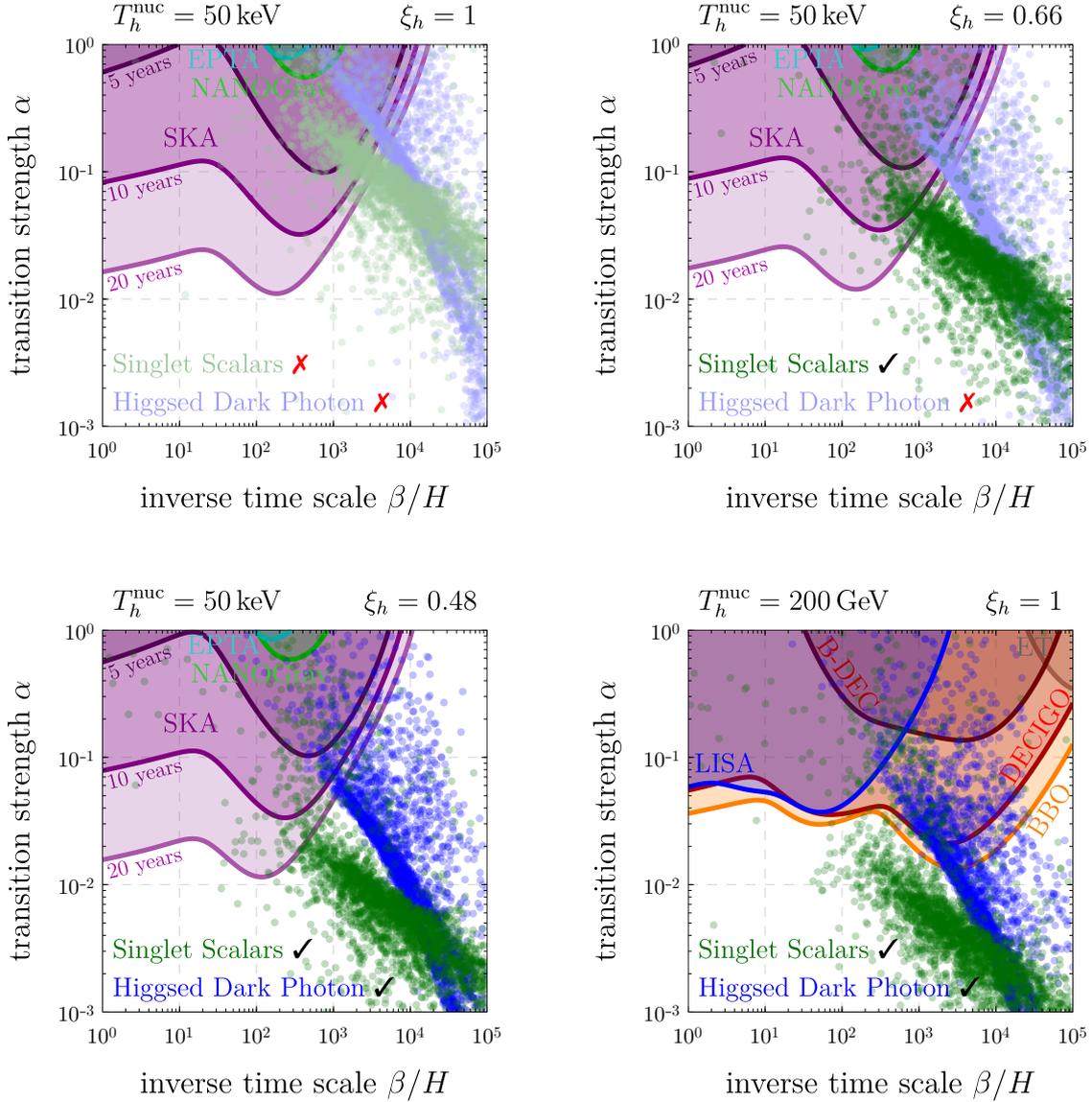

	\centering
	\includegraphics[width=0.495\textwidth, trim={0 0 0 0}, clip]
	{{{scatter/scatter_beta-alpha_T=10e-4.3_rT=1_alphainf=1alpha}}}
	\includegraphics[width=0.495\textwidth, trim={0 0 0 0}, clip]
	{{{scatter/scatter_beta-alpha_T=10e-4.3_rT=0.66_alphainf=1alpha}}}\\[0.8cm]
	\includegraphics[width=0.495\textwidth, trim={0 0 0 0}, clip]
	{{{scatter/scatter_beta-alpha_T=10e-4.3_rT=0.48_alphainf=1alpha}}}
	\includegraphics[width=0.495\textwidth, trim={0 0 0 0}, clip]
	{{{scatter/scatter_beta-alpha_T=10e2.3_rT=1_alphainf=1alpha}}}\\
	\caption{Ranges of gravitational wave parameters for the singlet scalar
		model (green points) and for the Higgsed dark photon model (blue points)
		from a random parameter scan.  For the singlet scalar model, we have
		scanned the parameter region $\log_{10}(\mu_A/v_S^0),\log_{10}(\kappa_{SA}/v_S^0),\log_{10}(\lambda_S),\log_{10}(\lambda_A) \in (-3,0)$,
		$\lambda_{SA} \in (0, 3)$, $\bar\kappa \in (0.7, 1.5)$. For the dark photon
		model, we have scanned over $\log_{10}(\lambda_S) \in (-4,-1)$ and
		$g_D \in (0, 1)$.  We compare to the expected sensitivities on
		various future gravitational wave observatories in the non-runaway
		regime ($\alpha < \alpha_\infty$), which is justified (approximately justified)
		for the Higgsed dark photon (singlet scalar model).
		The four panels correspond to different bubble nucleation temperatures,
		$\Thnuc$, and to different temperature ratios between the dark and visible
		sectors, $\temprat{h}$, as indicated in the various panels.
		The value $\temprat{h}=0.66$ ($\temprat{h}=0.48$) has been chosen such that a phase
		transition at $\Thnuc = \SI{50}{keV}$ satisfies the BBN and CMB$+H_0$ constraints
		for the singlet scalar model in the $\nu$\mbox{-}quilibration scenario
		(Higgsed dark photon model for a fully decoupled hidden sector). A black tick mark
		({\color{black}\ding{51}}) next to a model label means that the model is
		cosmologically allowed for the respective value of \temprat{h}.  A red cross
		({\color{red}\ding{55}}) indicates the model violates $\Neff$
		constraints.
	}
	\label{fig:model-scatter}
\end{figure}

Thus far we have shown only specific cuts through the parameter space of our
toy models. To better illustrate the full range of parameters that yields
observable signals we plot in \cref{fig:model-scatter} the results of random
parameter scans for both our toy models, overlaid with the sensitivities of
different future gravitational wave observatories.  For the singlet scalar
model, we have chosen the parameter ranges as follows: $\log_{10}(\mu_A/v_S^0),\log_{10}(\kappa_{SA}/v_S^0),\log_{10}(\lambda_S),\log_{10}(\lambda_A) \in (-3, 0)$,
$\lambda_{SA} \in (0, 3)$, $\bar\kappa \in (0.7, 1.5)$.  Note that the quartic
couplings $\lambda_S$ and $\lambda_A$ play only a negligible role for the
dynamics of the phase transition.  For the dark photon model we scan over
$\log_{10}(\lambda_S) \in (-4,-1)$, $g_D\in(0, 1)$. To produce the plots in
\cref{fig:model-scatter}, we have first determined the parameters $\alpha_h$
(i.e.\ the value of $\alpha$ assuming $\temprat{h} = 1$) and $\beta/H$ for
4\,000 random parameter points per model while keeping the value of $v_S^0$
fixed. Next, we have rescaled the resulting gravitational wave parameters such
that all random points have the same nucleation temperature $\Thnuc$ (indicated
in the plots). This rescaling affects $\alpha_h$ due to the temperature
dependence of $\gG$, and $\beta/H$ through the temperature dependence of the
nucleation condition.  Finally, we have rescaled $\alpha$ according to
\cref{eq:alpha,eq:rho-R} to the desired temperature ratio $\temprat{h}$ while
keeping \Thnuc fixed.

The resulting random points are displayed in \cref{fig:model-scatter} together
with the experimental sensitivities (shaded regions), i.e.\ the regions where
the signal-to-noise ratio in a given gravitational wave observatory exceeds the
detection threshold. For the determination of these regions, the temperature
ratio $\temprat{h}$ enters again as it alters the redshift of the signal
amplitude and peak frequency, see \cref{eq:full-redshift}.
\Cref{fig:model-scatter} reveals that the singlet scalar model features
transitions that tend to be weaker (smaller $\alpha$) compared to the Higgsed
dark photon model in most of the considered parameter space. Typically, there
exists a limit to how large the barrier between the true and false vacuum can
be without forcing the Universe into the true vacuum already at high
temperatures.  This places a lower limit on the size of $\beta/H$.
We also observe that the correlation in the $\alpha$--$\beta/H$ plane is more
pronounced for the dark photon model than for the singlet scalar model.  This
is related to the fact that in the dark photon model, the dynamics of the phase
transition has a significant dependence on only two of the particle physics
parameters ($\lambda_S$ and $g_D$), while in the singlet scalars model, there
is a dependence on five parameters.  At a nucleation temperature of $\Thnuc =
\SI{50}{keV}$, SKA will be sensitive to a significant portion of the parameter
space for both models, while EPTA and NANOGrav would only able to exclude a few
extreme points. At this scale, however, a temperature ratio $\temprat{h}<1$ is
required to ensure a consistent cosmology. For $\temprat{h}=0.66$ (upper-right
panel in \cref{fig:model-scatter}) the singlet scalar model is allowed in the
$\nu$\mbox{-}quilibration scenario if the less stringent CMB$+H_0$ constraint from
\cref{eq:Neff-constraint-CMB-H0} is applied. When assuming a fully decoupled
and even colder hidden sector, with $\temprat{h}=0.48$ (lower-left panel), both
models are allowed by the CMB+$H_0$ and BBN constraints.  At $\Thnuc = \SI{200}{GeV}$
(bottom-right panel), the latent heat released in the phase transition,
measured in units of to the total energy density of the Universe, is much
smaller than at $\Thnuc = \SI{50}{keV}$.  This results in values for $\alpha$
that are more than an order of magnitude smaller for otherwise identical model
parameters. Therefore, at $\Thnuc = \SI{200}{GeV}$, only the far-future
space-based interferometers DECIGO and BBO would be sensitive to the Higgsed
dark photon model, while LISA and B\mbox{-}DECIGO cover only very small portions of
its parameter spaces. Meanwhile, the transitions of the singlet scalar model
turn out to be mostly undetectable for $\Thnuc = \SI{200}{GeV}$.

Comparing the situation for a dark sector at the same temperature as the
photons ($\temprat{h} = 1$, upper-left of \cref{fig:model-scatter}) with
the situation for a cooler dark sector ($\temprat{h} = 0.66$ in the upper-right
panel and $\temprat{h} = 0.48$ in the lower-left panel),
we see that the associated rescaling of $\alpha \propto \temprat{h}^4$ moves many
parameter points outside the detectable window. Nevertheless, a
non-negligible fraction of points remains detectable for phase transitions at
low $\Thnuc$. 

We conclude that, in our toy models, a stochastic gravitational wave background
from a phase transition in the early Universe can be observed only if the
nucleation temperature is low (typically in the range \si{\keV}--\si{\MeV} for
the large $\beta/H$ exhibited) and the dark sector is significantly colder than
the visible sector (to avoid CMB and BBN constraints), or if we wait for far-future
observatories like DECIGO or BBO.

\section{Conclusions}
\label{sec:conclusions}

In summary, we have studied gravitational wave signals from first-order
cosmological phase transitions, in particular transitions occurring in
a secluded hidden sector.  We have paid special attention to phase transitions
at low (sub-MeV) temperatures.  Secluded hidden sectors at such low scales cannot
be explored using traditional approaches such as collider searches or direct
dark matter searches.  The main constraints instead arise from cosmological
measurements of the relativistic energy density, parameterized by $\Neff$.
Taking these constraints into account, we have first restricted the number of
hidden sector degrees of freedom as a function of the hidden sector temperature,
see \cref{fig:Neff-constraints}.

As expected, hidden sectors with non-trivial dynamics below $\sim \si{MeV}$ 
are only allowed if they are colder than the Standard Model sector by an
$\order{1}$ factor. We have then investigated the detectability of
gravitational wave signals from such a light hidden sector, focusing in
particular on the sensitivity of pulsar timing arrays, which are the most
relevant experiments for such low-frequency signals.  Noise curves and power
law-integrated sensitivity curves for a range of experiments are given in
\cref{fig:noise_and_power-law-int-sense}, and the data underlying this figure
is attached as ancillary material.  We have found that detectable signals are
only expected if the hidden sector is not too much colder than the Standard
Model sector, see for example the bottom row of \cref{fig:sensitive-regions}.
These requirements on the hidden sector temperature allow us to bound the
parameter range in which observable gravitational wave signals can occur from
both directions.  The results of which are shown in \cref{fig:max_hidden_dofs}.

We have also considered phase transitions at higher temperatures, where hardly
any cosmological constraints exist.  However, at high temperatures, the
gravitational wave energy density is typically much smaller compared to the
total energy density of the Universe, making the transition strength parameter
$\alpha$ small and the signal more challenging to detect.

We have finally corroborated the above statements by constructing two specific
toy models featuring phase transitions at sub-MeV temperatures: one with just
an extended scalar sector, and one with a Higgsed $U(1)'$ gauge symmetry in the
hidden sector.  Because of the above constraints on the number of hidden sector
degrees of freedom and on their temperature, we expect that many models with
observable low-temperature phase transitions reduce to one of our toy models,
or a model very similar to them, at low energies.  Thus, our toy models offer
interesting discovery opportunities for SKA in a range of gravitational wave
frequencies inaccessible to any other observatory.

\tocless{\section*{Acknowledgments}}

TO would like to thank Stefan Huber and Jose Miguel No for enlightening
discussions.  TO is also grateful to CERN for hospitality and support during
crucial stages of this project.  The authors have received funding from the
German Research Foundation (DFG) under Grant Nos.\ EXC-1098, FOR~2239, GRK~1581 and KO4820/4\mbox{-}1, and from the European Research Council (ERC) under the European
Union's Horizon 2020 research and innovation programme (grant agreement No.\
637506, ``$\nu$Directions'').
\vspace{5ex}

\appendix

\section{The Effective Potential}
\label{sec:appendix-effpot}

\subsection{General Formalism}
\label{sec:appendix-effpot-formalism}

A central ingredient in our analysis of the cosmological evolution of the toy
models from \cref{sec:toymodels} is the finite-temperature effective potential
\begin{align}
  V_\text{eff}(S,\Th) = V_\text{tree}(S)
                      + V_\text{CW}(S)
                      + V_T(S,\Th)
                      + V_\text{daisy}(S,\Th) \,,
\end{align}
evaluated at one-loop in the perturbative expansion.\footnote{Note that, in
  general, $V_\text{eff}$ depends on all scalar fields in the model. In our toy
  models however, we only care about the dependence on $S$: in the singlet
  scalar model, the auxiliary scalar $A$ never acquires a vev; in both models,
couplings between the hidden sector scalars and the SM Higgs field are assumed
to be tiny.} Here, $V_\text{tree}$ is the tree-level potential.  The one-loop
contribution splits up into a ultraviolet-divergent zero-temperature 
(Coleman--Weinberg)
part, $V_\text{CW}$, and a finite-temperature part, $V_T$.  We also include the
resummed contribution from ring diagrams, $V_\text{daisy}$.

The renormalized Coleman--Weinberg part is given by~\cite{Coleman:1973jx,
Quiros:1999jp, Laine:2016hma}
\begin{align}
  V_\text{CW}(S) &= \sum_i \frac{\eta_i n_i}{64 \pi^2} m_i^4(S)
                    \bb{\log\ba{\frac{m_i^2(S)}{\Lambda^2}} - C_i} + V_\text{ct}(S) \,,
\end{align}
where $i$ runs over all particle species with $S$-dependent mass, $n_i$ is the
number of degrees of freedom for each species, and
$\eta_i = +1$ ($-1$) for bosons (fermions).\footnote{Since we are working in
Landau gauge, the sums in this section are meant to include both the
  Goldstone bosons and the longitudinal gauge boson modes. Despite the
  Goldstone boson equivalence theorem this does not imply double counting, as
  demonstrated in~\cite{Delaunay:2007wb}.}
For the renormalization scale $\Lambda$, we choose the tree-level vev $v_S^0
\equiv
v_S(T=0)$. Ultraviolet divergences are canceled by counter-terms, while the
remaining finite parts of the counter-terms are subsumed in
$V_\text{ct}(S)$. The constant $C_i = 3/2$ ($5/6$) for scalars and fermions
(gauge bosons) is an artifact of dimensional regularization. We write the
finite part of the counter-terms as
\begin{align}
    V_\text{ct}(S) &= \frac{\delta\mu_S^2}{2} S^2 + \frac{\delta \kappa}{3}S^3 + \frac{\delta\lambda_S}{4} S^4
\end{align}
in the singlet scalar model from \cref{sec:singlet-scalar}, and as
\begin{align}
  V_\text{ct}(S) &= - \frac{\delta\mu_S^2}{2} S^2
                                  + \frac{\delta\lambda_S}{8} S^4
\end{align}
for the dark photon model from \cref{sec:dark-photon}.
We determine the counter-term couplings by imposing the renormalization conditions
\begin{align}
\begin{split}
  \be{\pd{V_\text{CW}(S)}{S}}{S=v_S^0}  &\overset{!}{=} 0 \,, \\
  \be{\pdd{V_\text{CW}(S)}{S}}{S=v_S^0} &\overset{!}{=} 0 \,,
\end{split}
\end{align}
which fix the zero-temperature vev and mass to their tree-level values.
For the singlet scalar model, we additionally want to fix the structure
of the minima, which is approximately achieved by requiring
\begin{align}
  V_\text{CW}(0) - V_\text{CW}(v_S^0) &\overset{!}{=} 0 \,.
\end{align}

The finite-temperature part of the one-loop potential evaluates
to~\cite{Dolan:1973qd, Quiros:1999jp, Laine:2016hma}
\begin{align}
  V_T(S,\Th) &= \sum_i \frac{\eta_i n_i \Th^4}{2\pi^2}
            \int_0^\infty \!\upd x \, x^2 \log \bb{1
                  - \eta_i \exp\ba{-\sqrt{x^2 + m_i^2(S)/\Th^2}} } \,.
\end{align}
At high temperature, this expression can be expanded as~\cite{Laine:2016hma}
\begin{align}
  \begin{split}
    V_T(S,\Th) &\simeq \Th^4 \sum_{\text{bosons}} n_i \bb{\frac{1}{24}\frac{m_i^2(S)}{\Th^2}
                          - \frac{1}{12\pi} \ba{\frac{m_i^2(S)}{\Th^2}}^{3/2}} \\
     & - \Th^4 \sum_{\text{fermions}} n_i \bb{\frac{1}{48}\frac{m_i^2(S)}{\Th^2}} \,,
  \end{split}
  \label{eq:VT-S}
\end{align}
where we have included only the leading field dependent pieces.

In addition to the one-loop contributions to $V_\text{eff}$,
we also consider the ring diagram (\enquote{daisy}) contributions,
which read~\cite{Carrington:1991hz, Quiros:1999jp, Delaunay:2007wb}
\begin{align}
  V_\text{daisy}(S,\Th) &= -\frac{\Th}{12\pi} \sum_\text{bosons} n_i
                          \bb{\ba{ m^2(S) + \Pi(\Th)}_i^{3/2}-\ba{m^2(S)}_i^{3/2}}
\end{align}
after resumming an infinite series of infrared-divergent diagrams.
Here, $\Pi(\Th)$ denotes the temperature dependent Debye mass,
which vanishes for transverse gauge boson modes~\cite{Carrington:1991hz}.
In the above formula, $(m^2(S) + \Pi(\Th))_i$ has to be interpreted as
the $i$-th eigenvalue of the full (tree-level + thermal) mass
matrix~\cite{Patel:2011th}. Here we note that the cubic terms
$\propto (m^2(S))^{3/2} \sim \abs{S}^3$ exactly cancel between the
high-temperature expansions of $V_T(S)$ and $V_\text{daisy}(S)$.
Coupling one or more gauge bosons to a scalar field is therefore
a possibility to generate a loop-induced barrier, rendering the phase transition
first-order. This is based on the fact that the cancellation of
cubic terms occurs only for longitudinal polarizations, but not for the
transverse polarizations. We utilize this barrier formed by the transverse
polarization to induce a first-order phase transition in the Higgsed dark photon
model of \cref{sec:dark-photon}.

\subsection{Model Details}

In the following, we list the scalar-field dependent particle masses,
thermal Debye masses, and tree-level minimization conditions for the
two toy models from \cref{sec:toymodels}.  These parameters need to be
plugged into the equations from \cref{sec:appendix-effpot-formalism} to compute
the effective potential $V_\text{eff}(S,\Th)$.

In the singlet scalar model, the tree-level minimization condition for the scalar
potential is
\begin{align}
  \mu^2_S = -\bb{\kappa + \lambda_S v_S^0} v_S^0 \,.
\end{align}
The field-dependent scalar masses of $S$ and $A$ are
\begin{align}
  m^2_S(S,A) &= \mu_S^2 + 2\kappa S + 3\lambda_S S^2 + \lambda_{SA} A^2 \,,
                                                  \label{eq:mS2} \\
  m^2_A(S,A) &= \mu_A^2 + 3\lambda_A A^2 + 2\kappa_{SA} S +\lambda_{SA} S^2
                                                  \label{eq:mA2} \,,
\end{align}
and their Debye masses are
\begin{align}
  \Pi_S(\Th) &= \bb{\frac{\lambda_S}{4} + \frac{\lambda_{SA}}{12}} \Th^2 \,, \\
  \Pi_A(\Th) &= \bb{\frac{\lambda_A}{4} + \frac{\lambda_{SA}}{12}} \Th^2 \,.
\end{align}

In the Higgsed dark photon model, the scalar potential is at its tree-level
minimum when
\begin{align}
  \mu^2_S = \frac{\lambda_S}{2}(v^0_S)^2 \,.
  \label{eq:min-cond-dark-photon}
\end{align}
The field-dependent masses of the two
scalar degrees of freedom and of the dark photon are
\begin{align}
  m^2_{\phi_S}(S)   &= -\mu_S^2+\frac{3}{2}\lambda_S S^2 \,, \\
  m^2_{\sigma_S}(S) &= -\mu_S^2+\frac{1}{2}\lambda_S S^2 \,, \\
  m^2_{A'}(S)       &= g_D^2 S^2 \,.
\end{align}
Note that, at the minimum of the potential ($S = v_S^0$), $\phi_S$
corresponds to the massive scalar degree of freedom, while
$\sigma_S$ is the Goldstone mode, as can be easily seen from
\cref{eq:min-cond-dark-photon}. The Debye masses are
\begin{align}
  \Pi_{S}(\Th)  &= \bb{\frac{\lambda_S}{6} + \frac{g_D^2}{4}} \Th^2 \,, \\
  \Pi_{A'}(\Th) &= \frac{g_D^2}{3} \Th^2 \,.
\end{align}

\section{Sensitivity Curves}
\label{sec:appendix-sensitivity}

In the following, we explain in detail how the experimental sensitivity curves
discussed in \cref{sec:sensitivities} (see for instance
\cref{fig:noise_and_power-law-int-sense}) have been obtained.

\subsection{Signal-to-Noise Ratio}
\label{sec:appendix-SNR}

Consider a system of $N$ gravitational wave detectors with output
\begin{align}
  s_i(t) = h_i(t) + n_i(t) \,,\hspace{3eM} i=1,\ldots,N\,,
\end{align}
where $h_i(t)$ is the strain induced by the gravitational wave signal we are
looking for and $n_i(t)$ is the noise in detector $i$. (Both quantities are assumed
to be already convoluted with the detector response function, see e.g.\
Ref.~\cite{Thrane:2013oya}.)
One can then define a (pair-wise) cross-correlated signal observed over a long
time interval \Tobs by
\begin{align}
  S_{ij} \equiv \int\limits_{-\Tobs/2}^{\Tobs/2} \!\!\upd t
                \int\limits_{-\Tobs/2}^{\Tobs/2} \!\!\upd t' \,
                s_i(t) \, s_j(t') \, Q_{ij}(t-t') \,,
  \label{eq:SNR-Sij}
\end{align}
where $Q(t)$ is a filter function. The filter function will be chosen such that it
maximizes the signal-to-noise ratio
\begin{align}
  \SNR_{ij} = \frac{\ev{S_{ij}}}{\sqrt{\ev{S_{ij}^2} - \ev{S_{ij}}^2}} \,,
  \label{eq:SNR-1}
\end{align}
where $\ev{\cdot}$ denotes an ensemble average, realized either by integrating
over different spatial regions or by integrating over many observations, each of
duration \Tobs.
In the following, it is convenient to work with the Fourier transforms of
$s_i(t)$, $h_i(t)$, $n_i(t)$, and $Q_{ij}(t)$, which are defined as usual by
\begin{align}
  \tilde{Q}_{ij}(f) &= \int_0^\infty \!\upd t \, Q_{ij}(t) \, e^{2\pi i f t} \,,
                                                          \label{eq:Qij-fourier} \\
  \tilde{s}_i(f)    &= \int_0^\infty \!\upd t \, s_i(t) \, e^{2\pi i f t} \,,
                                                          \label{eq:si-fourier}
\end{align}
and similar expressions for $\tilde{h}_i(f)$ and $\tilde{n}_i(f)$.
Assuming that the gravitational wave background and the noise are Gaussian and
stationary, and that the noise levels in the individual detectors are statistically
independent, one can define the power spectral densities
$S_h(f)$ and $P_{ni}(f)$ via~\cite{Allen:1997ad,Thrane:2013oya,Caprini:2018mtu}
\begin{align}
  \ev{\tilde{h}_i(f) \tilde{h}_j^*(f')} &=
    \frac{1}{2} \delta(f - f') \, \Gamma_{ij}(f) \, S_{h}(f)
\intertext{and}
  \ev{\tilde{n}_i(f) \tilde{n}_j^*(f')} &=
    \frac{1}{2} \delta(f - f') \,\delta_{ij} \, P_{ni}(f) \,.
\end{align}
Here, $\Gamma_{ij}(f)$ is called the overlap reduction function. It encodes
the sky-averaged and polarization-averaged detector response to an incoming
gravitational wave, taking into account the reduction in sensitivity due
to different locations and orientations of the two detectors $i$ and $j$. Note
that, for the noise, we use a quantity $P_{ni}(f)$ defined directly in terms
of the detector response, whereas for the signal, it is more convenient to work
with $S_h(f)$, which has detector effects ($\Gamma_{ij}(f)$) factored out.
$S_h(f)$ is related to the fractional cosmological gravitational wave energy
density spectrum $\Omega_\text{GW}(f)$ (see \cref{sec:gw-spectrum}) via
\begin{align}
  S_h(f) = \frac{3 H_0^2}{2 \pi^2} \frac{\Omega_\text{GW}(f)}{f^3} \,.
  \label{eq:Sh-OmegaGW}
\end{align}
With these definitions, and with the optimally chosen
\begin{align}
  \tilde{Q}_{ij}(f) \propto \frac{\Gamma_{ij}(f) S_h(f)}{P_{ni}(f) P_{nj}(f)}
\end{align}
(see for instance \cite{Caprini:2018mtu}),
the signal-to-noise-ratio from \cref{eq:SNR-1} can be rewritten as
\begin{align}
  \SNR^2 &= \sum\limits_{i,\,j>i} \SNR_{ij}^2
          = 2 \,\Tobs \left(\frac{3 H_0^2}{2\pi^2}\right)^2
            \sum\limits_{j>i} \, \int\limits_{f_\text{min}}^{f_\text{max}} \!\upd f \,
            \frac{\Gamma^2_{ij}(f) \, \big[\Omega_{\text{GW}}(f)\big]^2}{f^6 \, P_{ni}(f) \, P_{nj}(f)} \,,
  \label{eq:SNR-2}
\end{align}
Here, $f_\text{min}$ and $f_\text{max}$ are the bounds of the frequency region
to which the detectors are sensitive. Frequencies outside this region do not
contribute to the signal-to-noise ratio.  It is finally convenient to define
the noise energy density spectrum of the detector network as
\begin{align}
  \Omega_\text{eff}(f) \equiv
    \frac{2 \pi^2 f^3}{3 H_0^2}
    \bigg[ \sum\limits_{j>i} \frac{\Gamma_{ij}^{2}(f)}{P_{ni}(f) P_{nj}
    (f)} \bigg]^{-1/2} \,.
  \label{eq:Sensitivity-EffectiveNoise}
\end{align}
With this definition, we can bring the signal-to-noise ratio to its final
form~\cite{Thrane:2013oya, Caprini:2018mtu}
\begin{align}
  \SNR^2 &= 2 \,\Tobs \int\limits_{f_\text{min}}^{f_\text{max}} \!\upd f \,
            \left[ \frac{\hhOmega{\text{GW}}(f)}{\hhOmega{\text{eff}}(f)}
            \right]^2\,,
         &\text{(cross-correlated between several detectors)}\,.
  \label{eq:Sensitivity-SNR}
\end{align}
This is just \cref{eq:Sensitivity-crosscorrelatedSNR}.

Given the effective noise curve $\hhOmega{\text{eff}}(f)$
and the threshold value $\SNR_\text{thr}$,
we can now evaluate whether a given stochastic gravitational wave background
can be detected.\footnote{Note, however, that in obtaining \cref{eq:Sensitivity-SNR},
the filter function $\tilde{Q}_{ij}(f)$ has been optimized for the
expected gravitational wave signal, requiring a priori knowledge of this
signal. Thus, when dealing with real data, the analysis would have to be
repeated for each type of signal to be tested against. This is typically not
feasible, therefore, the sensitivity obtained in this way should be regarded as
only an estimate.}

\subsection{Pulsar Timing Arrays}
\label{sec:PTA}

\begin{table}
  \begin{tabular}{lC{2cm}C{2.2cm}C{2cm}C{2cm}C{2cm}l}
    \toprule
    Experiment & $N_p$ & \Tobs & $\delta t$ & $\sigma$ & $\rho_\text{thr}$ & Reference \\
    \hline
    EPTA & $6$ & $\SI{8}{}-\SI{18}{years}$ & $\SI{10}{days}$ & $\SI{0.1}{}-\SI{1.7}{\micro s}$ & $1.19$ & \cite{Lentati:2015qwp, Desvignes:2016yex} \\
    NANOGrav & $34$ & $\SI{4}{}-\SI{11}{years}$ & $\SI{7}{}-\SI{30}{days}$ & $\SI{0.1}{}-\SI{3.7}{\micro s}$ & $0.697$ & \cite{Arzoumanian:2018saf} \\
    SKA  & $1000$ & $5$, $10$, $\SI{20}{years}$ & $\SI{14}{days}$ & $\SI{100}{ns}$ & 4& \cite{Bull:2018lat,Janssen:2014dka} (assumptions) \\
    \botrule
  \end{tabular}
  \caption{Number of pulsars $N_p$, observation time \Tobs, observation
    interval $\delta t$, timing uncertainty $\sigma$ and signal-to-noise
    detection threshold $\rho_\text{thr}$ for the PTAs
    considered in our study.}
  \label{tab:pta-parameters}
\end{table}

PTA limits on a stochastic gravitational wave background are usually quoted in terms of the
minimal detectable amplitude of the characteristic strain
\begin{align}
  h_c(f) \equiv \sqrt{f \, S_h} \,,
  \label{eq:characteristic-strain}
\end{align}
which we parameterize as a simple power-law, i.e.
\begin{align}
  h_c(f) = A_a \bigg( \frac{f}{\bar f} \bigg)^a \,.
  \label{eq:PTA-strain}
\end{align}
Here, $A_a$ is the strain amplitude at an arbitrary reference frequency
which we choose as $\bar f = \SI{1}{yr^{-1}}$.
The background in a PTA is typically assumed to be generated by unresolved
supermassive black hole binary (SMBHB) systems, in which case $a = -2/3$,
or by cosmic strings with $a=-1$~\cite{Lentati:2015qwp, Arzoumanian:2018saf,
Caprini:2018mtu}. The corresponding gravitational wave power spectrum is
related to the strain by~\cite{Thrane:2013oya,Moore:2014lga}
\begin{align}
  \Omega_\text{GW}(f) = \frac{2 \pi^2}{3 H_0^2} f^2 h_c^2(f)
                      \equiv \Omega_b \left(\frac{f}{\bar f}\right)^b\,,
  \label{eq:PTA-h2Omega}
\end{align}
with $b=2+2 a$ (see also \cref{eq:Sensitivity-PLSpectra}).

We consider here current constraints from the observation of pulsars over a
time span of 18~years by EPTA~\cite{Lentati:2015qwp} and over 11~years from
NANOGrav~\cite{Arzoumanian:2018saf}, as well as prospective limits from
SKA~\cite{Janssen:2014dka}.  EPTA and NANOGrav present their results as limits
on the gravitational wave amplitude as a function of frequency for a spectrum
with freely-varying spectral index obtained using a Bayesian approach.  We use
these curves as an estimate for $\hhOmega{\text{eff}}$ and take the
signal-to-noise
threshold $\SNR_\text{thr}$ equal to the signal-to-noise that saturates the limits on the
SMBHB background ($a = -2/3$). These limits are $A_{-2/3} \leq 1.45 \times
10^{-15}$ for NANOGrav~\cite{Arzoumanian:2018saf} and
$\hhOmega{\text{SMBHB}}(2.8\,\si{nHz}) \leq 1.1\times 10^{-9}$ for
EPTA~\cite{Lentati:2015qwp}, which leads to $\SNR_\text{thr} = 1.19$ for EPTA
with $\Tobs = \SI{18}{yrs}$ and $\SNR_\text{thr} = 0.697$ for NANOGrav with
$\Tobs = \SI{11}{yrs}$.

\begin{figure}
  \begin{center}
    \includegraphics[width=0.48\textwidth]{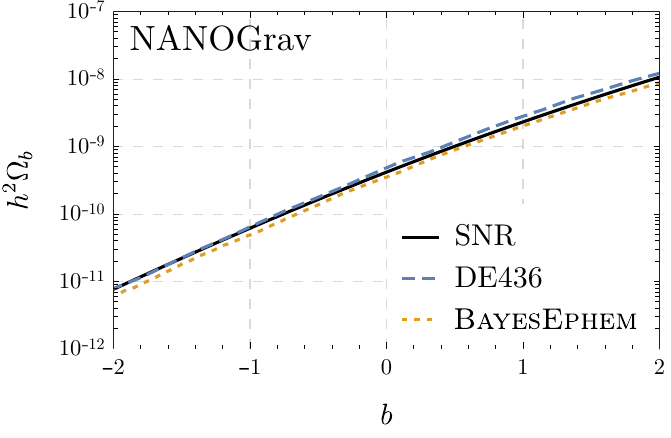}
    \hfill
    \includegraphics[width=0.48\textwidth]{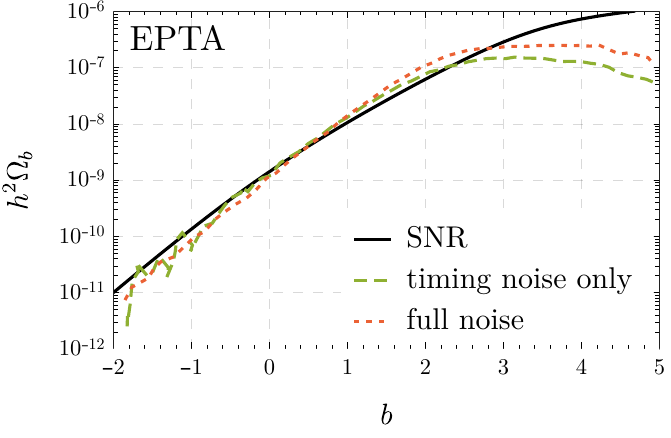}
  \end{center}
  \caption{95\,\%~CL upper limit on the energy density of a stochastic
    gravitational wave background with spectral slope $b$ from NANOGrav (left)
    and EPTA (right).  In both plots, the black solid line is our estimate
    based on \cref{eq:Sensitivity-SNR} and the signal-to-noise ratio thresholds
    given in the text.  In the NANOGrav plot, the dashed blue and dotted orange lines
    are taken from Fig.~2 of Ref.~\cite{Arzoumanian:2018saf}. The dashed blue
    line corresponds to a specific set of ephemerides (DE436), while the dotted orange
    one is marginalized over ephemeris uncertainties.
    The dashed green and dotted red line in the
    EPTA plot correspond to different noise models, see Fig.~14 of
    Ref.~\cite{Lentati:2015qwp}.
  }
  \label{fig:PTA-SpectralLimit}
\end{figure}

As a cross-check we have used these $\SNR_\text{thr}$ values in
\cref{eq:Sensitivity-SNR} to compute the limits on the gravitational
wave amplitude at $f = \SI{1}{yr^{-1}}$ as a function of the spectral slope.
In \cref{fig:PTA-SpectralLimit}, we compare these limits to the bounds derived
by the collaborations. We see that for spectral slopes $-2 \leq b \leq 2$,
our limits agree very well with the ones provided by the collaborations.
For larger $b$, the official EPTA limits are somewhat stronger than our
estimates. (NANOGrav does not show limits in this range.)

\begin{figure}
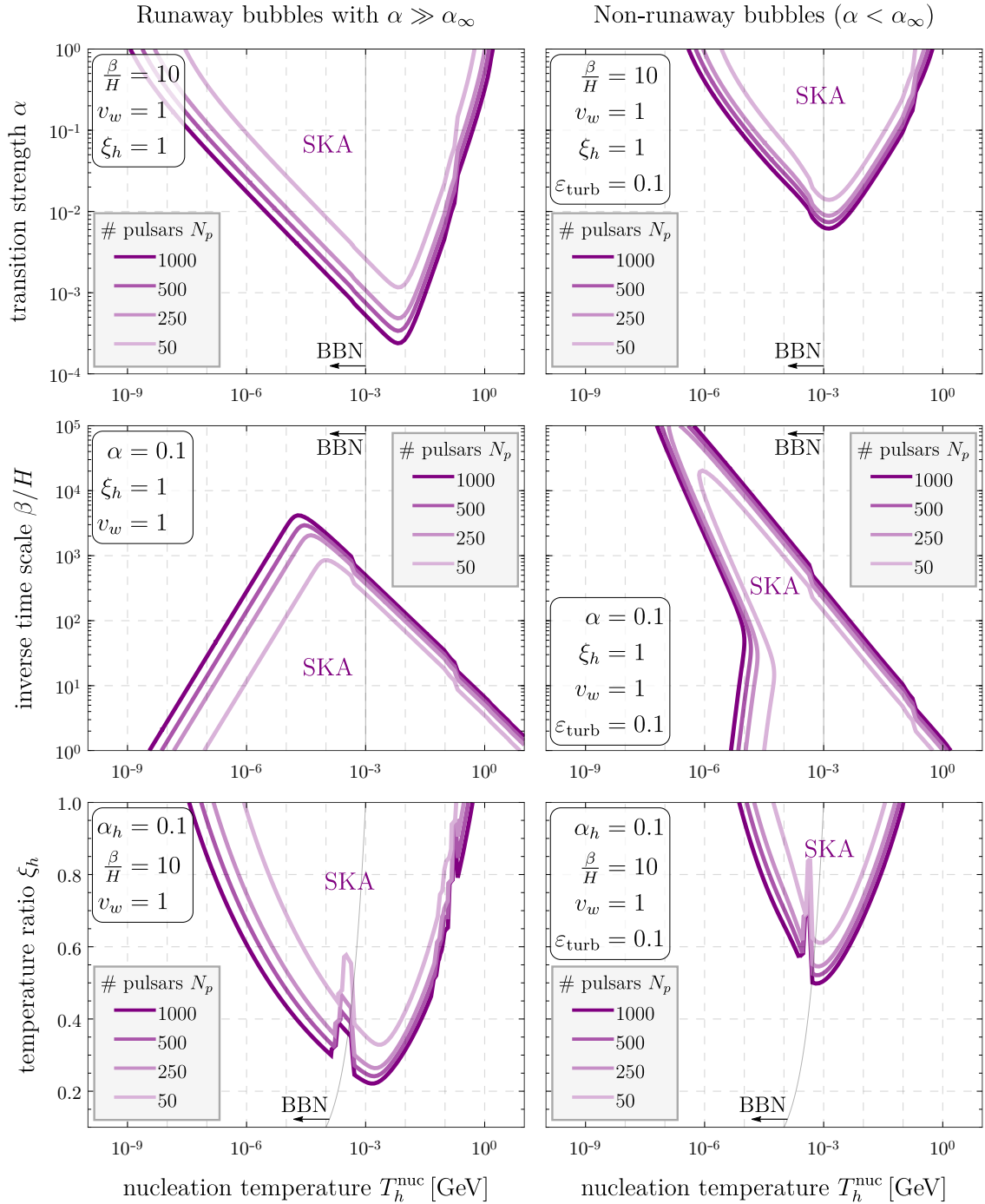

  \centering
  \includegraphics[width=\textwidth, trim={1.65cm 1.2cm 1.65cm 0.4cm}, clip]
  {{{SKA/T-alpha_beta=10}}}\\
  \includegraphics[width=\textwidth, trim={1.65cm 1.2cm 1.65cm 1cm}, clip]
  {{{SKA/T-beta_alpha=0.1}}}\\
  \includegraphics[width=\textwidth, trim={1.65cm 0.5cm 1.65cm 1cm}, clip]
  {{{SKA/T-rT_alpha=0.1_beta=10}}}
  \caption{Anticipated sensitivity to hidden sector phase transitions with the
    SKA telescope, assuming alternate values for the number of timed pulsars
   $N_p$. In the left-hand panels
    we assume runaway bubbles, while the right-hand panels show the case of
    non-runaway bubbles. We show the sensitivities as a function of the
    hidden sector temperature at which bubble nucleation occurs, $\Thnuc$,
    versus the transition strength $\alpha$ (top), the inverse time scale
    $\beta/H$ (middle), and the temperature ratio between the hidden and visible
    sector $\temprat{h}$ (bottom). In all panels, we have assumed $\DOFh \ll
    \gG$ in calculating the redshifting of gravitational wave spectra.
    Note that in the bottom panel, we fix
    $\alpha_h$ (the value of $\alpha$ at $\temprat{h} = 1$) instead of
    $\alpha$ to explicitly show the $\temprat{h}$-dependence of $\alpha$ for
    fixed values of \Thnuc. Note that the translation of $\alpha_h$ to the
    physical $\alpha$ also relies upon the assumption $\DOFh \ll
    \gG$. The discontinuities originate from the step-function
   approximation for $\gG$.
  }
  \label{fig:VariedPulsarNumber}
\end{figure}

For the sensitivity of SKA, we work in the weak signal limit where the noise is
the dominant contribution to the detector output (this is particularly good
assumption when timing a large number of pulsars).  In addition we assume that
the timing-residual noise of each pulsar is white, Gaussian and uncorrelated.
The noise power spectral density can then be written as~\cite{Thrane:2013oya,
Moore:2014eua}
\begin{align}
  P_n(f) = 2 \sigma^2 \, \delta t \,,
\end{align}
where $\delta t$ is the inverse of the pulsar's cadence (which specifies how
frequently the pulsar's timing residual is measured) and $\sigma$ is the
residual root-mean-square (rms) error on each of these measurements.
We assume white noise, i.e.\ there is no frequency dependence in $\sigma$.
For an array of $N_p$ pulsars randomly distributed over the sky and with equal timing
cadence and noise, the effective noise power spectral density then
becomes~\cite{Moore:2014eua}
\begin{align}
  S_\text{eff}(f) &=
    \Bigg( \sum_{i=1}^{N_p} \sum_{j=i+1}^{N_p}
           \frac{\Gamma_{ij}^2(f)}{P_n^2(f)} \Bigg)^{-\frac{1}{2}} \,.
  \label{eq:PTA-Seff-1}
\end{align}
(Note that, by considering the ratio $\Gamma_{ij}^2 / P_n^2(f)$ here, we
deconvolute the detector response, i.e.\ $S_\text{eff}$ can be directly
compared to the signal power spectral density $S_h(f)$.)
For our numerical calculations, we approximate $\Gamma_{ij}(f)$ as
\cite{Moore:2014eua}
\begin{align}
  \Gamma_{ij}(f) = \frac{1}{12 \pi^2 f^2} \frac{1}{4\sqrt{3}} \,,
\end{align}
which leads to
\begin{align}
  S_\text{eff}(f) &=
    96 \sqrt{3} \pi^2 f^2 \, \sigma^2 \delta t \, \sqrt{ \frac{2}{N_p(N_p -1)}} \,.
  \label{eq:PTA-Seff-2}
\end{align}
In analogy to 
\cref{eq:characteristic-strain,eq:PTA-h2Omega},
$S_\text{eff}$ can be related to a characteristic strain $h_\text{eff}^2 \equiv
f S_\text{eff}(f)$~\cite{Moore:2014eua,Moore:2014lga}, and thus to
an effective energy density spectrum (normalized to the critical density
of the Universe),
\begin{align}
  \Omega_\text{eff}(f) = \sqrt{\frac{2}{N_p (N_p - 1)}}
                         \frac{64 \sqrt{3} \pi^4 \sigma^2 \delta t}{H_0^2} f^5\,.
  \label{eq:PTA-h2OmegaEff}
\end{align}
This quantity can then be used in \cref{eq:Sensitivity-SNR} to determine the
signal-to-noise ratio. The integration limits in \cref{eq:Sensitivity-SNR}
should be taken as $f_\text{min} = 1/\Tobs$ and $f_\text{max} =
1/\delta t$~\cite{Moore:2014lga}.
Using \cref{eq:PTA-h2OmegaEff,eq:Sensitivity-SNR}, it is straightforward to
rescale the noise and sensitivity curves in \cref{fig:noise_and_power-law-int-sense}
to the actual experimental parameters of SKA once these are known. Nevertheless,
in \cref{fig:VariedPulsarNumber} we show the changes in sensitivity to the
gravitational wave parameters when varying the number of pulsars timed.
Note that reducing
the number of pulsars can affect the validity of the weak signal approximation.
Timing around 50 pulsars over an observation run of 20 years yields already
percent
level
deviations between the signal-to-noise defined using \cref{eq:PTA-h2OmegaEff}, and the full
result (relaxing the assumption of a
weak signal) in Eq.~(19) of Ref.~\cite{Siemens:2013zla}.

As a cross-check we have considered the hypothetical SKA campaign taken from 
Ref.~\cite{Janssen:2014dka}, which is based on $N_p =
50$ pulsars timed once per week ($\delta t = 7\,\si{days}$) over a time span of
$\Tobs = \SI{10}{\years}$.  The residual noise has been assumed to be $\sigma
=\SI{100}{\nano\s}$, and the detection threshold has been taken as
$\SNR_\text{thr}=4$.  
Plugging these numbers
into \cref{eq:PTA-h2OmegaEff} for the noise and \cref{eq:PTA-h2Omega} for the
signal (with $b = 2/3$ for a stochastic gravitational wave signal due to
SMBHBs), and using the result in \cref{eq:Sensitivity-SNR} for the
signal-to-noise ratio, we find that the minimal detectable energy density for a
SMBHB signal is $\hhOmega{\text{SMBHB}}(\bar f) = \SI{2E-12}{}$ at a frequency of
$\bar f = \SI{1}{yr^{-1}}$.  Converted into a characteristic strain this gives
$A_{-2/3} = \SI{5.7E-17}{}$, which agrees with the estimate $A_{-2/3} \sim
\SI{E-17}{}$--$\SI{E-16}{}$ from Ref.~\cite{Janssen:2014dka}.  (Note that in
the analyses presented in the main part of the paper, we consider a more
optimistic SKA configuration, as given in \cref{tab:pta-parameters} based upon
the feasibility study of Ref.~\cite{Bull:2018lat}.)

\subsection{Space-Based Interferometers}

In this work we consider the projected sensitivities for the LISA
experiment~\cite{Audley:2017drz}, which is planned to be launched in
2034~\cite{Caprini:2018mtu}, as well as the proposed successor experiments
BBO~\cite{Crowder:2005nr}, B\mbox{-}DECIGO~\cite{Sathyaprakash:2012jk}, and
DECIGO~\cite{Seto:2001qf}.

While DECIGO and BBO will consist of networks of gravitational wave detectors,
LISA and B\mbox{-}DECIGO are single-detector observatories.
Hence, the cross-correlation analysis that has led to 
\cref{eq:Sensitivity-SNR} is not directly applicable to LISA and B\mbox{-}DECIGO.
Instead, we need to use the auto-correlated signal-to-noise
\begin{align}
  \SNR^2 &= \Tobs \int\limits_{f_\text{min}}^{f_\text{max}} \! \upd f \,
            \left[ \frac{\hhOmega{\text{GW}}(f)}{\hhOmega{\text{eff}}(f)}
            \right]^2\,,
         & \text{(single detector)}\,.
  \label{eq:Sensitivity-SNRLISA}
\end{align}
This expression differs from \cref{eq:Sensitivity-SNR} by just a factor of two
and is thus equal to the squared cross-correlated signal-to-noise for a pair of detectors
divided by two, since we only use one detector.

For LISA we assume a mission duration of $\Tobs = \SI{4}{yrs}$~\cite{Audley:2017drz}
and an signal-to-noise detection threshold of $\SNR_\text{thr} = 10$~\cite{Caprini:2015zlo}. 
We use the noise strain power spectral density quoted in Ref.~\cite{Cornish:2018dyw}
to be
\begin{align}
  S_\text{eff}^\text{LISA}(f) = \frac{10}{3 L^2}
           \bigg( P_\text{OMS}(f)
                + 2 \bigg[1 + \cos^2\bigg( \frac{f}{f_\ast} \bigg) \bigg] 
                      \frac{P_\text{acc}(f)}{(2\pi f)^4} 
                     \bigg)
                    \bigg[ 1 + \frac{6}{10} \bigg( \frac{f}{f_\ast} \bigg)^2
                    \bigg]
                + S_c(f) \,,
  \label{eq:Seff-LISA}
\end{align}
where $L = 2.5 \times 10^9\,\si{m}$ is LISA's interferometer arm length and
$f_\ast = c/(2\pi L)$ is called the transfer frequency where $c$ is the
speed of light.
The instrument noise is composed of the optical metrology noise
\begin{align}
  P_\text{OMS}(f) = (\SI{1.5e-11}{m})^2 \,
                    \bigg[ 1 + \bigg( \frac{\SI{2}{mHz}}{f} \bigg)^4 \bigg] \, \si{Hz^{-1}}
\end{align}
and the test mass acceleration noise
\begin{align}
  P_\text{acc}(f) = (\SI{3e-15}{m\,sec^{-2}})^2 \,
                    \bigg[ 1 + \bigg( \frac{\SI{0.4}{mHz}}{f} \bigg)^2 \bigg]
                    \bigg[ 1 + \bigg( \frac{f}{\SI{8}{mHz}} \bigg)^4 \bigg] \, \si{Hz^{-1}}\,.
\end{align}
The confusion noise from unresolved galactic binaries after 4 years
is
\begin{multline}
  S_c(f) = 9 \times 10^{-45} \, f^{-7/3}\
           \exp\Bigg[ - \bigg(\frac{f}{\si{Hz}}\bigg)^{0.138}
                      - 221 \bigg(\frac{f}{\si{Hz}}\bigg)
                            \sin\bigg[521 \bigg( \frac{f}{\si{Hz}}\bigg) \bigg] \Bigg] \\
        \times  \bigg\{ 1 + \tanh \bigg[1680 \bigg(0.0013 - 
                              \bigg( \frac{f}{\si{Hz}} \bigg) \bigg) \bigg] \bigg\} \,\si{Hz^{-1}}\ .
\end{multline}

For B\mbox{-}DECIGO (the scaled-down predecessor of DECIGO, with a planned launch in
the late 2020's~\cite{Sato:2017dkf}), the effective noise strain power spectral
density is~\cite{Isoyama:2018rjb}
\begin{multline}
  S_\text{eff}^\text{B-DECIGO}(f) =
  2.020 \times 10^{-45}  \bigg[ 1 + 1.584 \times 10^{-2}
                                            \bigg(\frac{f}{\si{Hz}} \bigg)^{-4} \\
                                  + 1.584 \times 10^{-3}
                                            \bigg(\frac{f}{\si{Hz}} \bigg)^2
                         \bigg] \, \si{Hz^{-1}} \,.
  \label{eq:Seff-B-DECIGO}
\end{multline}
(Note that we include an additional factor $5$ compared to Ref.~\cite{Isoyama:2018rjb}
to account for sky-averaging.)
The frequency range for B\mbox{-}DECIGO is $f_\text{min} = \SI{0.01}{Hz}$,
$f_\text{max} = \SI{100}{Hz}$, the detection threshold is $\SNR_\text{thr} = 8$,
and the assumed observation time is 4 years.

\begin{table}
	\begin{tabular}{lC{4cm}C{1.75cm}C{1cm}l}
		\toprule
		Experiment & Frequency range & $\SNR_\text{thr}$ & $\hhOmega{\text{eff}}$ & Comment \\
		\hline
		LISA & \hfill$\SI{e-5}{}-\SI{1}{Hz}$\hfill\llap{\cite{Audley:2017drz}} & \hfill$10$\hfill\llap{\cite{Caprini:2015zlo}} & \cite{Cornish:2018dyw} & \cref{eq:Sensitivity-SNRLISA} for SNR \\
		B\mbox{-}DECIGO & \hfill$\SI{e-2}{}-\SI{e2}{Hz}$\hfill\llap{\cite{Isoyama:2018rjb}} & \hfill$8$\hfill\llap{\cite{Isoyama:2018rjb}} & \cite{Isoyama:2018rjb} & $\times5$ for sky-average and \cref{eq:Sensitivity-SNRLISA} for SNR \\
		DECIGO & \hfill$\SI{e-3}{}-\SI{e2}{Hz}$\hfill\llap{\cite{Yagi:2013du}} & $10$ & \cite{Yagi:2013du} & $\times5$ for sky-average \\
		BBO & \hfill$\SI{e-3}{}-\SI{e2}{Hz}$\hfill\llap{\cite{Yagi:2011yu}} & $10$ &\cite{Yagi:2011yu} & $\times5$ for sky-average \\
		ET & \hfill$\SI{1}{}-\SI{e4}{Hz}$\hfill\llap{\cite{Sathyaprakash:2012jk}} & \hfill$5$\hfill\llap{\cite{Sathyaprakash:2012jk}} & \cite{Sathyaprakash:2012jk} & \cref{eq:Sensitivity-SNRLISA} for SNR \\
		\botrule
	\end{tabular}
	\caption{Parameters and assumptions made for future space-based and
		ground-based interferometers. We assume $\Tobs=\SI{4}{\years}$ for all
		space-based experiments, as proposed for LISA~\cite{Audley:2017drz} and
		B\mbox{-}DECIGO~\cite{Isoyama:2018rjb}, and $\Tobs=\SI{5}{\years}$ for
		ET~\cite{Sathyaprakash:2012jk}.}
	\label{tab:experiment-parameters}
\end{table}

The noise curves for the far-future projects BBO~\cite{Crowder:2005nr} and
DECIGO \cite{Sato:2017dkf} can be parameterized as~\cite{Yagi:2011yu,Yagi:2013du}
\begin{align}
  S_\text{eff}^\text{BBO}(f) =
    5 \times \min \bb{ \frac{S_n^\text{inst}(f)}{\exp\ba{-\kappa \,T \,\upd N/\upd f}}, \,
                       S_n^\text{inst}(f) + S_n^\text{gal}(f) \mathcal{F}(f)}
                + S_n^\text{ex-gal}(f)\,,
  \label{eq:Seff-BBO}
\end{align}
where $\kappa=4.5$, and $\upd N / \upd f = 2 \times (f / \si{Hz})^{-11/3} \,
\si{Hz^{-1}}$ is the spectral number density of galactic white dwarf binaries.
The non-sky-averaged instrumental noise curves are
\begin{multline}
  S_n^\text{inst, DECIGO}(f)
    = 5.3 \times 10^{-48} \times \Bigg(
        \bigg[ 1 + \bigg( \frac{f}{\fp} \bigg)^2 \bigg] 
    + 2.3\times 10^{-7} \bigg( \frac{f}{f_p} \bigg)^{-4} \frac{1}{1 + (f/\fp)^2} \\
    + 2.6\times 10^{-8} \bigg( \frac{f}{f_p} \bigg)^{-4} 
  \Bigg) \, \si{Hz^{-1}}
\end{multline}
with $\fp = \SI{7.36}{Hz}$ for DECIGO, and
\begin{align}
  S_n^\text{inst,BBO}(f)= \Bigg[
                 1.8 \times 10^{-49} \bigg(\frac{f}{\si{Hz}} \bigg)^2
               + 2.9 \times 10^{-49}
               + 9.2 \times 10^{-52} \bigg(\frac{f}{\si{Hz}} \bigg)^{-4} 
        \Bigg]\,\si{Hz^{-1}}
\end{align}
for BBO.
The confusion noise from galactic ($S_n^\text{gal}$)
and extra-galactic ($S_n^
\text{ex-gal}$) white dwarf binaries is given by
\begin{align}
  S_n^\text{gal}    &= 2.1 \times 10^{-45} \ba{\frac{f}{\si{Hz}}}^{-7/3} \, \si{Hz^{-1}} \,, \\
  S_n^\text{ex-gal} &= 4.2 \times 10^{-47} \ba{\frac{f}{\si{Hz}}}^{-7/3} \, \si{Hz^{-1}} \,.
\end{align}
As for LISA, we use an observation time of $\Tobs = \SI{4}{yrs}$ and a detection
threshold of $\SNR_\text{thr} = 10$.  The frequency range covered by DECIGO and BBO
will be from \SI{0.1}{Hz} to \SI{10}{Hz}. Our parameter choices for future
gravitational wave interferometers are also
summarized in \cref{tab:experiment-parameters}.

\subsection{Earth-Based Interferometers}

Due to seismic noise the sensitivity of current ground-based gravitational wave
observatories is not sufficient to constrain stochastic backgrounds generated
by cosmological phase transitions.  However, the next-generation of detectors
is going to provide a significant improvement in sensitivity.

We hence include sensitivity projections for the third-generation observatory
ET~\cite{Yagi:2011yu,Yagi:2013du}. We employ the ET-D
noise projections~\cite{Hild:2010id}, assuming an observation period of 5 years
and considering a stochastic background as detectable if it produces an signal-to-noise of
5 in a single ET detector \cite{Sathyaprakash:2012jk}.

\bibliographystyle{JHEP}
\tocless\bibliography{refs}

\end{document}